\documentclass[12pt]{article}
\usepackage[pdftex]{graphicx}
\usepackage[colorlinks=true,linkcolor=blue,citecolor=blue]{hyperref}
\usepackage{url}
\usepackage{lineno}
\usepackage{geometry}
\usepackage{titlesec}
\usepackage{authblk}
\usepackage{subcaption} 
\usepackage[export]{adjustbox}
\usepackage{caption}
\usepackage{tikz}
\usetikzlibrary{tikzmark}
\usetikzlibrary {arrows.meta}
\usepackage{wrapfig}
\usepackage{amsmath} 
\usepackage[detect-all=true,separate-uncertainty=true]{siunitx}
\usepackage[capitalise]{cleveref}
\usepackage[T1]{fontenc}
\usepackage{booktabs}         
\usepackage{tabularx}         
\usepackage{array}            
\usepackage[backend=biber,style=numeric,sorting=none,giveninits=true,maxbibnames=15]{biblatex}
\newcommand{\npmt}{\hat{N}_{\mathrm{trig}}^{\mathrm{PMT}}}
\newcommand{\npmtnohat}{N_{\mathrm{trig}}^{\mathrm{PMT}}}
\newcommand{\dd}{\mathrm{d}}
\newcolumntype{Y}{>{\raggedright\arraybackslash}X}
\newcolumntype{C}{>{\centering\arraybackslash}X}
\titlespacing{\section}{0pt}{2ex}{1ex}
\titlespacing{\subsection}{0pt}{1.5ex}{0.5ex}

\title{Extreme High-Energy Neutrinos: IceCube vs.\ KM3NeT}
\author[1]{Lu Lu\thanks{lu.lu@icecube.wisc.edu}}
\author[1]{Tianlu Yuan\thanks{tyuan@icecube.wisc.edu}}
\affil[1]{Department of Physics and Wisconsin IceCube Particle Astrophysics Center, University of Wisconsin, Madison, WI 53706, USA}

\date{}  

\begin{document}

\maketitle
\vspace{-1cm}
\begin{abstract}
We review the state of the art in the detection of extreme high-energy neutrinos, focusing on the IceCube and KM3NeT neutrino telescopes. IceCube, operating deep in Antarctic ice, and KM3NeT, a new array in the Mediterranean Sea, employ distinct designs to capture Cherenkov light from neutrino interactions. We examine their detector architectures, readout and reconstruction performance for PeV-scale and higher-energy neutrinos. Recent candidate events above \SI{5}{\peta \eV} are highlighted. These include a \SI{\sim{120}}{\peta \eV} muon track observed by KM3NeT in 2023, and IceCube’s highest-energy detections, which comprise several-PeV showers and tracks. We outline current approaches to neutrino energy reconstruction and explore scenarios that might explain the apparent differences in observed event characteristics. Finally, we summarize future prospects for extreme-energy neutrino observations and their implications for astrophysical source populations and cosmogenic neutrinos.
\end{abstract}

\textbf{Keywords:} neutrino telescopes, IceCube, KM3NeT, Cherenkov


\section{Introduction}
\label{sec:intro}
Extensive air showers produced by ultra-high-energy cosmic rays (UHECR) were first observed by Pierre Auger in 1938~\cite{Auger:1939sh}, 
following the development of coincidence circuits that enabled their detection~\cite{1930Natur.125..636R}. 
With the subsequent discovery of the cosmic microwave background (CMB)~\cite{Penzias:1965wn,Dicke:1965zz}, 
Greisen, Zatsepin, and Kuzmin predicted that neutrinos would be produced 
when UHE protons interact with CMB photons via $\Delta$-resonance~\cite{Greisen:1966jv,Zatsepin:1966jv}, resulting in charged pions that then decay into neutrinos and muons, as well as neutrons that may undergo beta decay~\cite{Gaisser1990}. These are now known as GZK neutrinos, or cosmogenic neutrinos, and the full production chain is shown in \cref{eq1}.
\begin{equation}
\vcenter{\hbox{%
    \begin{tikzpicture}[remember picture,every node/.style={outer sep=2pt},->]
        \node {$p^+ + \gamma_\mathrm{CMB} \rightarrow \Delta^+ \rightarrow \tikzmarknode{L1-1}{n} + \tikzmarknode{L1-2}\pi^+$};
        
        \draw (L1-1) |-++ (.5,-1.5) node[right] {$\tikzmarknode{L4-1}{\mathrm{e}}^- + \tikzmarknode{L4-2}{p}^+ + \bar{\nu}_\mathrm{e}$};

        \draw (L1-2) |-++ (.2,-.5) node[right] {$\tikzmarknode{L2-1}{\mu}^+ + \nu_\mu$};
        
        \draw (L2-1) |-++ (.2,-.5) node[right] {$\mathrm{e}^+ + \nu_{\mathrm{e}} + \bar{\nu}_\mu $};
        
    \end{tikzpicture}
}}
\label{eq1}
\end{equation}

Determining the composition of UHECRs remains a major challenge~\cite{Kampert:2012mx,Anchordoqui:2018qom}, 
largely because the relevant particle interactions occur at center-of-mass energies several orders of magnitude above those accessible at the Large Hadron Collider (LHC)~\cite{Evans:2008zzb,Ostapchenko:2010vb}. 
One example of this challenge is the ``muon puzzle''---the observed excess of muons in extensive air showers compared to predictions from LHC-tuned Monte Carlo simulations~\cite{PierreAuger:2014ucz,PierreAuger:2016nfk}. 
In contrast, the electromagnetic (EM) component of air showers is less sensitive to hadronic interaction uncertainties~\cite{Matthews:2005sd}, 
and thus provides a more robust observable for composition studies~\cite{Kampert:2012mx}.

At the highest energies, measurements of the EM component of extensive air showers---particularly via the depth of shower maximum ($X_{\text{max}}$) observed by the Pierre Auger Observatory's fluorescence telescopes~\cite{PierreAuger:2010ymv,PierreAuger:2014sui}---indicate a \emph{mixed mass composition} rather than a composition dominated by protons~\cite{PierreAuger:2016use,Unger:2015laa}. 
This finding has significant implications for GZK neutrino production~\cite{Kotera:2011cp,Ahlers:2012rz}, as the resulting neutrino flux is highly sensitive to the primary cosmic ray composition (with protons yielding more neutrinos than heavier nuclei)~\cite{Kotera:2011cp}, the redshift evolution of the sources~\cite{Berezinsky:2002nc}, the injected spectral index at the source~\cite{Aloisio:2013hya}, and the maximum acceleration energy ($E_{\text{max}}$) achievable in the source environment~\cite{Hillas:1984ijl}.

To detect extremely high-energy (EHE) neutrinos effectively, a detector must have a large effective area and target volume~\cite{Alvarez-Muñiz:2017rp,Halzen:2010yj}, 
due to the low flux and interaction cross-section of such particles~\cite{Connolly:2011vc}. 
For example, the Pierre Auger Observatory uses the atmosphere as part of the detector and features a 3000\,km$^2$ surface array of water-Cherenkov detectors capable of identifying Earth-skimming $\nu_\tau$ events~\cite{PierreAuger:2007vvh,PierreAuger:2015ihf}. 
The IceCube Neutrino Observatory, located at the geographic South Pole, consists of a cubic-kilometer array of photomultiplier tubes (PMT) attached to cabling, or ``strings'', embedded in Antarctic ice and detects Cherenkov light from high-energy neutrino interactions~\cite{Aartsen:2016nxy,Halzen:2010yj}.

In recent years, several large-scale neutrino telescopes have been constructed or proposed~\cite{Katz:2011ke,Decoene:2023beq}. 
These include the KM3NeT experiment in the Mediterranean Sea~\cite{KM3Net:2016zxf}, the Baikal-GVD detector in Lake Baikal~\cite{Baikal-GVD:2018isr}, and the Pacific Ocean Neutrino Experiment (P-ONE) off the coast of Canada~\cite{P-ONE:2020ljt}. 
Additionally, three proposed experiments---HUNT, NEON, and TRIDENT---aim to operate in the South China Sea~\cite{Huang:2023mzt,Zhang:2024slv,TRIDENT:2022hql}.

The field of EHE neutrino detection had, for decades, produced only upper limits on fluxes, with no confirmed candidate events. This changed in February 2023, when the KM3NeT Collaboration—while operating in an incomplete configuration with only 21 DUs—recorded the EHE neutrino candidate event KM3-230213A~\cite{KM3NeT:2025npi}. The event, detected by the ARCA detector, corresponds to a muon with reconstructed energy of \SI{120(110:60)}{\peta \eV}. Assuming an $E^{-2}$ spectrum, this corresponds to a median neutrino energy of $\sim{\SI{220}{PeV}}$. This marks the most energetic neutrino candidate ever observed in Cherenkov detectors.

\begin{figure}[htbp]
  \centering
  \includegraphics[width=0.9\textwidth]{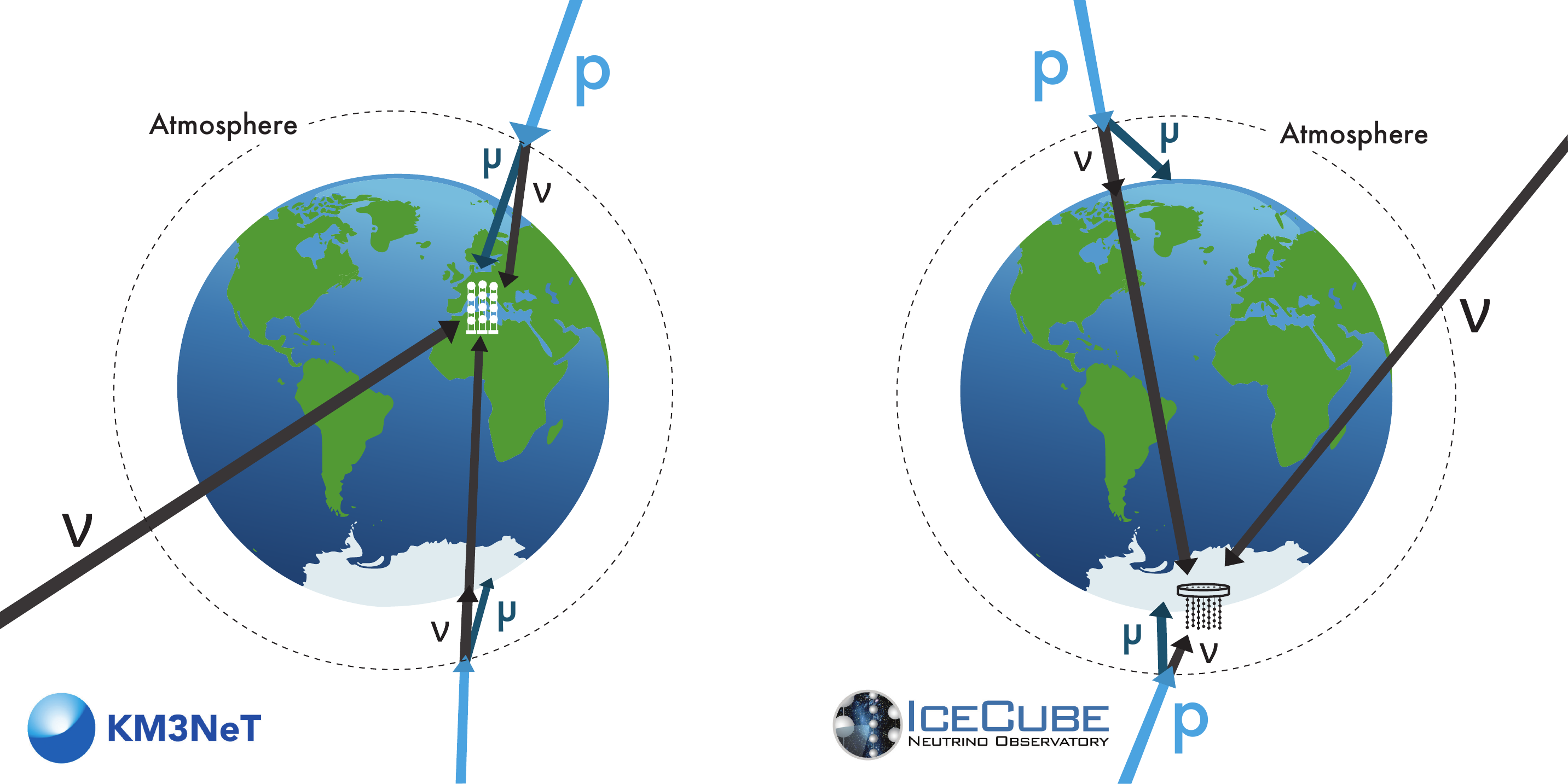}
  \caption{Illustration of KM3NeT (left) and IceCube (right) with their locations on Earth. While atmospheric muons are absorbed by the Earth, neutrinos can traverse it unattenuated until much higher energies. The two observatories provide complementary measurements, covering different regions of the sky. (credit: B. Sherman)}
  \label{fig:fig1}
\end{figure}

In this review, we cover the state of the art in EHE neutrino detection~\cite{Alvarez-Muñiz:2017rp,Halzen:2010yj,Katz:2011ke}, 
focusing on the reconstruction strategies and recent results from current-generation observatories~\cite{IceCube:2020acn,KM3Net:2016zxf}, 
with particular emphasis on KM3NeT and IceCube (illustrated in \cref{fig:fig1}). The rest of this section provides an overview of the IceCube and KM3NeT neutrino observatories, and a discussion of the diffuse neutrino flux measurements by those two experiments to-date. \Cref{sec:detect} reviews how both observatories aim to detect and reconstruct EHE neutrinos, and gives a summary of neutrino candidates above \SI{5}{\peta \eV}. \Cref{sec:disc} details the inferred neutrino energy of KM3-230213A, and interprets it in the context of non-observations by other observatories at that energy. Finally, prospects for future detectors are outlined in \cref{sec:future}. We note that the distinction between EHE and UHE neutrinos is somewhat arbitrary, and different terminology has been used by different groups when discussing the same event or analysis. Here, we adopt the terminology EHE when referring to neutrino events or analyses focused on energies above \SI{5}{\peta \eV}.

\subsection{IceCube and KM3NeT Neutrino Observatories}

Above tens of GeV, the neutrino--nucleon deep-inelastic-scattering (DIS) cross section is the dominant interaction process and grows approximately linearly with neutrino energy up to the TeV scale, with a slower, sub-linear power-law dependence at higher energies determined by parton distribution functions at small Bjorken-$x$~\cite{Gandhi:1998ri,Connolly:2011vc}. 
At neutrino energies of $\mathcal{O}(\SI{10}{\peta \eV})$, roughly half of neutrinos traversing the Earth at a zenith angle of about $10^\circ$ below the horizon are absorbed due to their interactions within the Earth~\cite{Learned:2000sw,Gaisser:1994yf}. Consequently, EHE neutrinos predominantly arrive from near-horizontal or down-going directions~\cite{Halzen:2006mq}. This leads to complementary sky coverage between major neutrino observatories: IceCube, located at the Geographic South Pole~\cite{Aartsen:2016nxy}, primarily observes EHE neutrinos from the Southern celestial hemisphere, 
while KM3NeT/ARCA is situated in the Northern Hemisphere offshore from Sicily~\cite{KM3Net:2016zxf} and capable of detecting such events from the Northern sky.

A major challenge in the detection of down-going astrophysical neutrinos is the suppression of background from atmospheric muons, which are produced in cosmic-ray interactions in the atmosphere. These muons can penetrate many kilometers into dense media and dominate the event rate from down-going directions. To mitigate this background, the PMTs are placed at depth to take advantage of the natural shielding provided by overlying material. The slant depth that muons must traverse before reaching the detector is referred to as the \emph{overburden}, and plays a critical role in the reduction of the atmospheric muon flux~\cite{Honda:2004yz}. In media such as water or ice, a \SI{10}{\peta \eV} muon can propagate over distances of about 15--20~km~\cite{Lipari1993_MuonProp}, losing energies via ionization, bremsstrahlung, pair production, and photonuclear interactions~\cite{Groom:2001kq}. A sufficiently large overburden, therefore, aids in achieving the background suppression required for EHE neutrino searches.

\begin{figure}[htbp]
  \centering
  \includegraphics[width=0.3\textwidth]{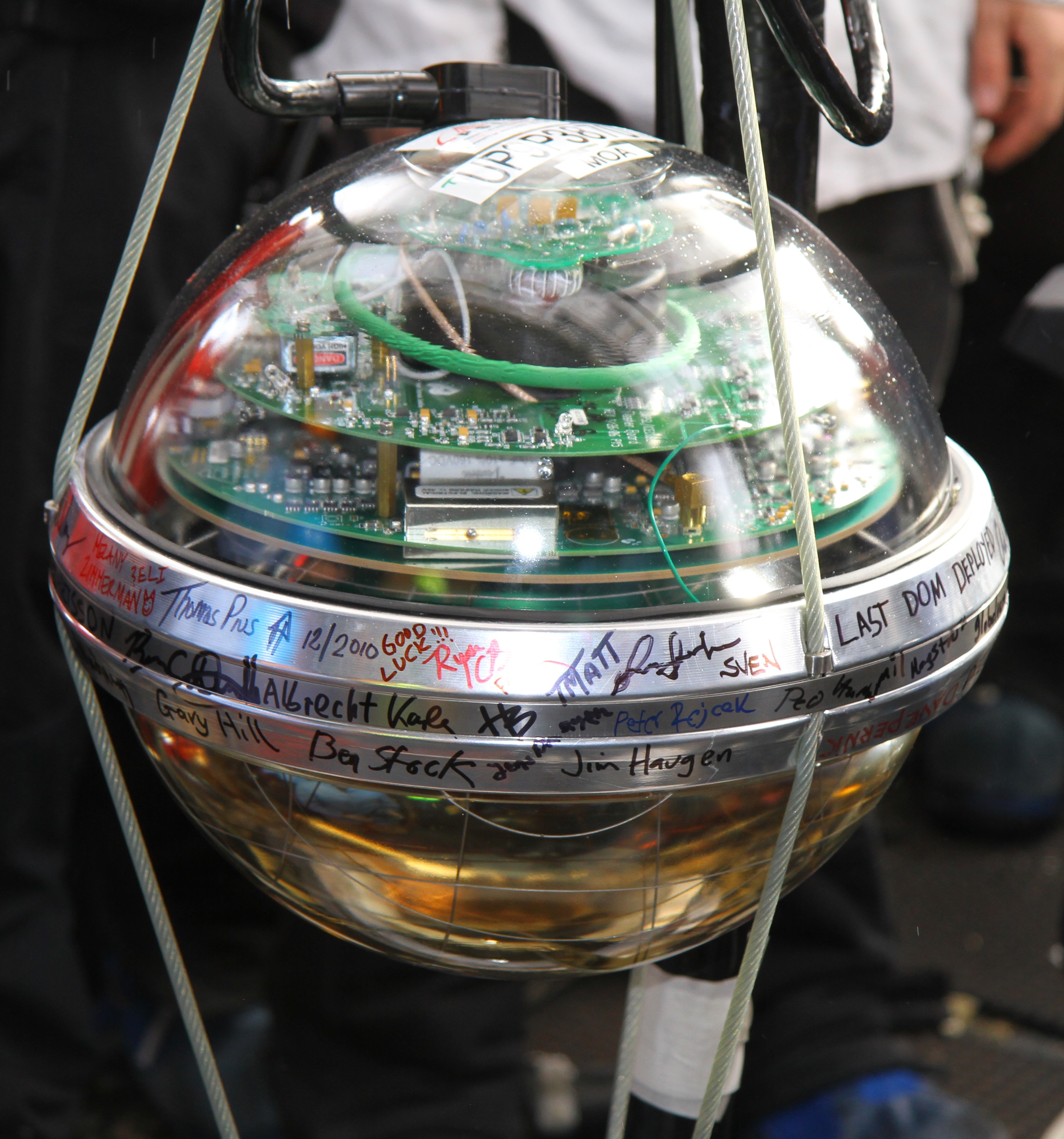}
  \includegraphics[width=0.295\textwidth]{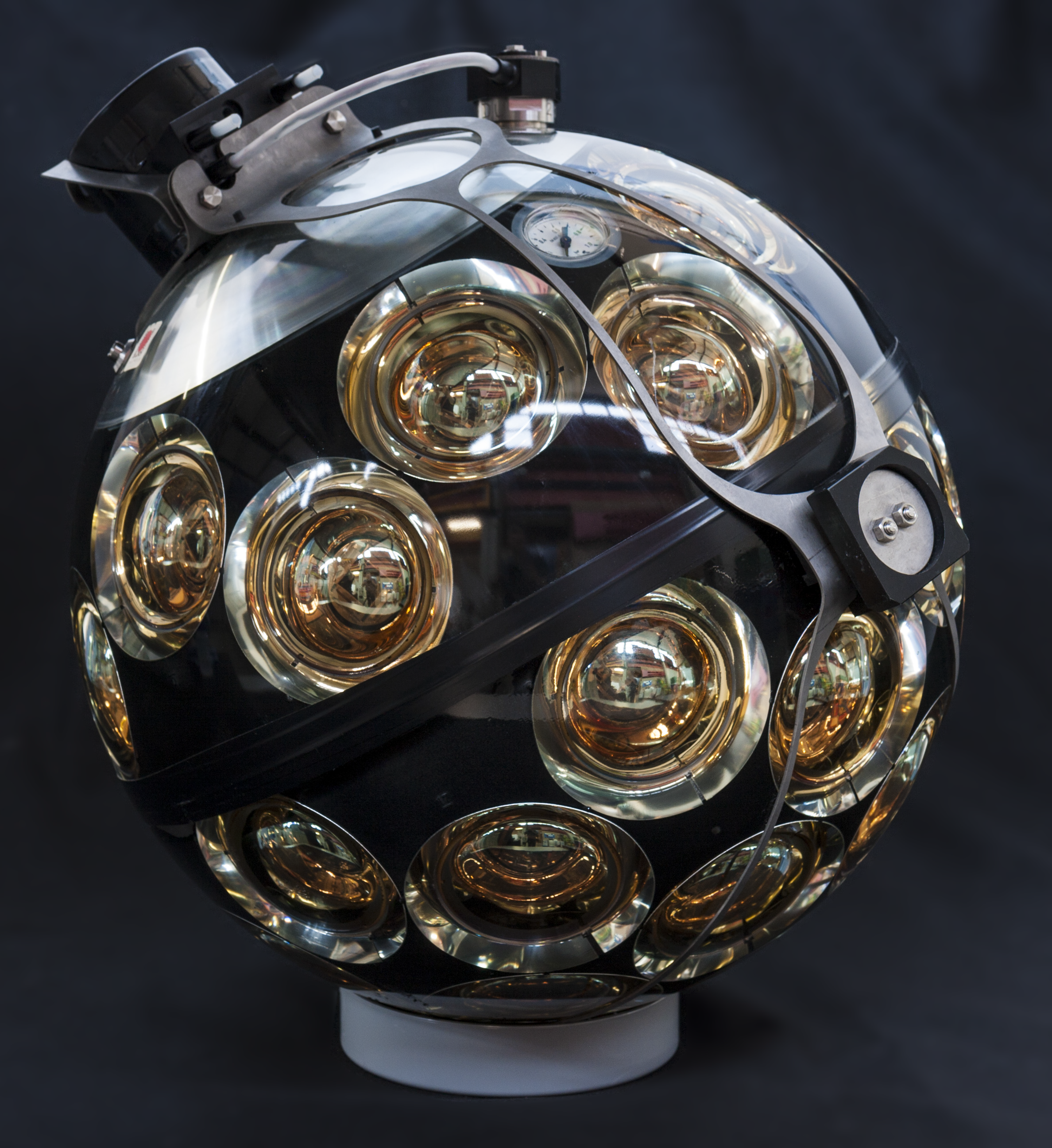}
  \caption{The left panel shows the IceCube optical module with a single, ten-inch, downward-facing PMT~\cite{Aartsen:2016nxy}. The right panel shows the KM3NeT optical module with 31 three-inch PMTs~\cite{KM3NeT:2022pnv}.}
  \label{fig:dom_comparison}
\end{figure}

Both IceCube and KM3NeT attempt to detect Cherenkov radiation emitted by secondary charged particles that are produced in neutrino DIS interactions. 
IceCube is embedded in the South Pole ice at depths between 1450~m and 2450~m~\cite{Aartsen:2016nxy}, 
while KM3NeT/ARCA is anchored on the seabed of the Mediterranean Sea offshore from Sicily, at a depth of about 3500~m~\cite{KM3Net:2016zxf,KM3NeT:2018wnd}. Both detectors rely on PMTs~\cite{HamamatsuPMTspec} housed in pressure-resistant glass spheres to detect Cherenkov photons. In IceCube, each Digital Optical Module (DOM) contains a single 10-inch downward-facing Hamamatsu PMT~\cite{Aartsen:2016nxy}. 
In contrast, each KM3NeT DOM contains 31 three-inch PMTs pointing in different directions~\cite{KM3NeT:2022pnv} in order to collect more information for directional reconstruction. \Cref{fig:dom_comparison} shows the IceCube and KM3NeT DOMs in the left and right panels, respectively.

The IceCube detector comprises 86 vertical strings (including DeepCore)~\cite{Aartsen:2016nxy}, 
with standard strings spaced $\sim{125}$~m apart and vertical DOM spacing of about 17~m~\cite{Aartsen:2016nxy}. 
Each string hosts 60 DOMs. The central region of IceCube contains the DeepCore sub-array, which consists of 8 densely instrumented strings~\cite{IceCube:2011ucd}. 
DeepCore enables a lower energy threshold in the few-GeV range, facilitating studies of neutrino oscillations~\cite{IceCube:2017lak} and atmospheric neutrinos~\cite{IceCube:2015mgt}.

IceCube was completed in December 2010 and has been continuously taking data with a detector uptime exceeding 99\%~\cite{Aartsen:2016nxy,IceCubeStatusReports}. 
A small fraction of DOMs have gradually become non-operational since deployment, 
but the overall detector performance remains stable and well-characterized.

The KM3NeT observatory is designed as a dual-site infrastructure comprising two detectors: Astroparticle Research with Cosmics in the Abyss (ARCA) for high-energy neutrino astronomy and Oscillation Research with Cosmics in the Abyss (ORCA) for low-energy neutrino oscillation studies~\cite{KM3Net:2016zxf}. 
ARCA consists of sparsely instrumented vertical detection units (DUs) with horizontal spacing of $\sim90$~m and vertical DOM spacing of 36~m. 
Each ARCA DU hosts 18 DOMs along a 700~m vertical line, optimized for detecting TeV--PeV neutrinos. 
ORCA, located offshore from Toulon, France, is a more densely instrumented array with string spacing of $\sim20$~m and vertical DOM spacing of 9~m, optimized for neutrinos in the few-GeV range. 
The projected instrumented volume of ARCA is on the order of 1~km$^3$, comparable to IceCube, while ORCA covers a smaller volume of several megatons, sufficient for low-energy precision measurements.

\begin{figure}[htbp]
  \centering
  \includegraphics[width=0.85\textwidth]{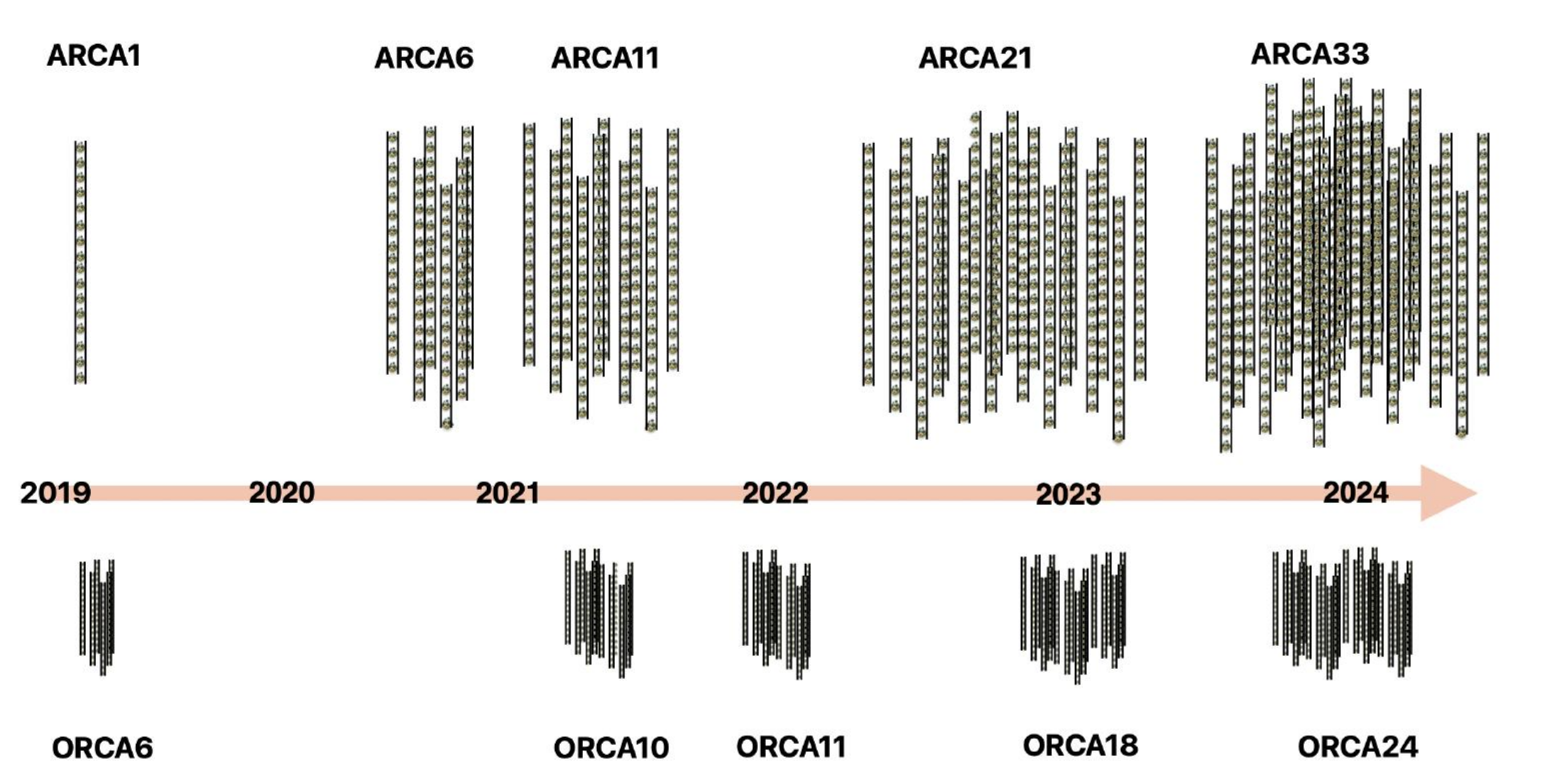}
  \caption{
    Deployment timeline of the KM3NeT neutrino telescope~\cite{Dornic2025_KM3NeTStatus}. 
    The smaller-volume ORCA detector is optimized for atmospheric neutrino oscillation measurements, 
    while the larger ARCA array is designed to detect astrophysical neutrinos at TeV energies and above. 
    Upon completion, ARCA is projected to comprise 230 detection units and instrument a volume of 1~km$^3$.
  }
  \label{fig:km3_timeline}
\end{figure}

A timeline of the KM3NeT construction phases~\cite{Dornic2025_KM3NeTStatus} is shown in \cref{fig:km3_timeline}. 
KM3NeT/ARCA is being deployed in stages, and as of 2025 has made steady progress toward its full configuration of 230 DUs. 
The EHE neutrino candidate event KM3NeT-230223A was detected during the ARCA21 configuration, which included 21 operational DUs~\cite{KM3NeT:2025npi}.

\begin{figure}[htbp]
  \centering
  \includegraphics[width=0.49\linewidth]{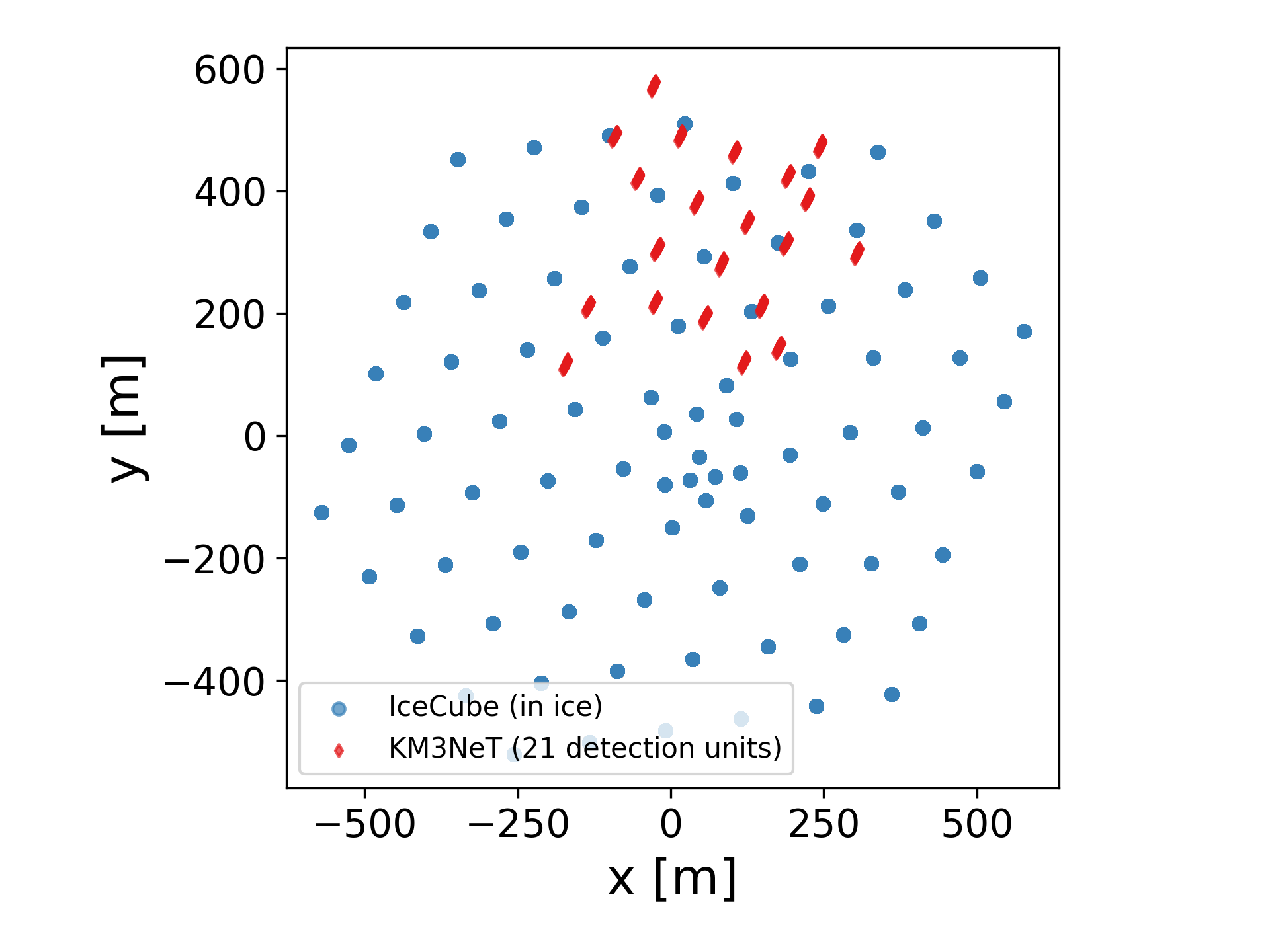}
  \includegraphics[width=0.49\linewidth]{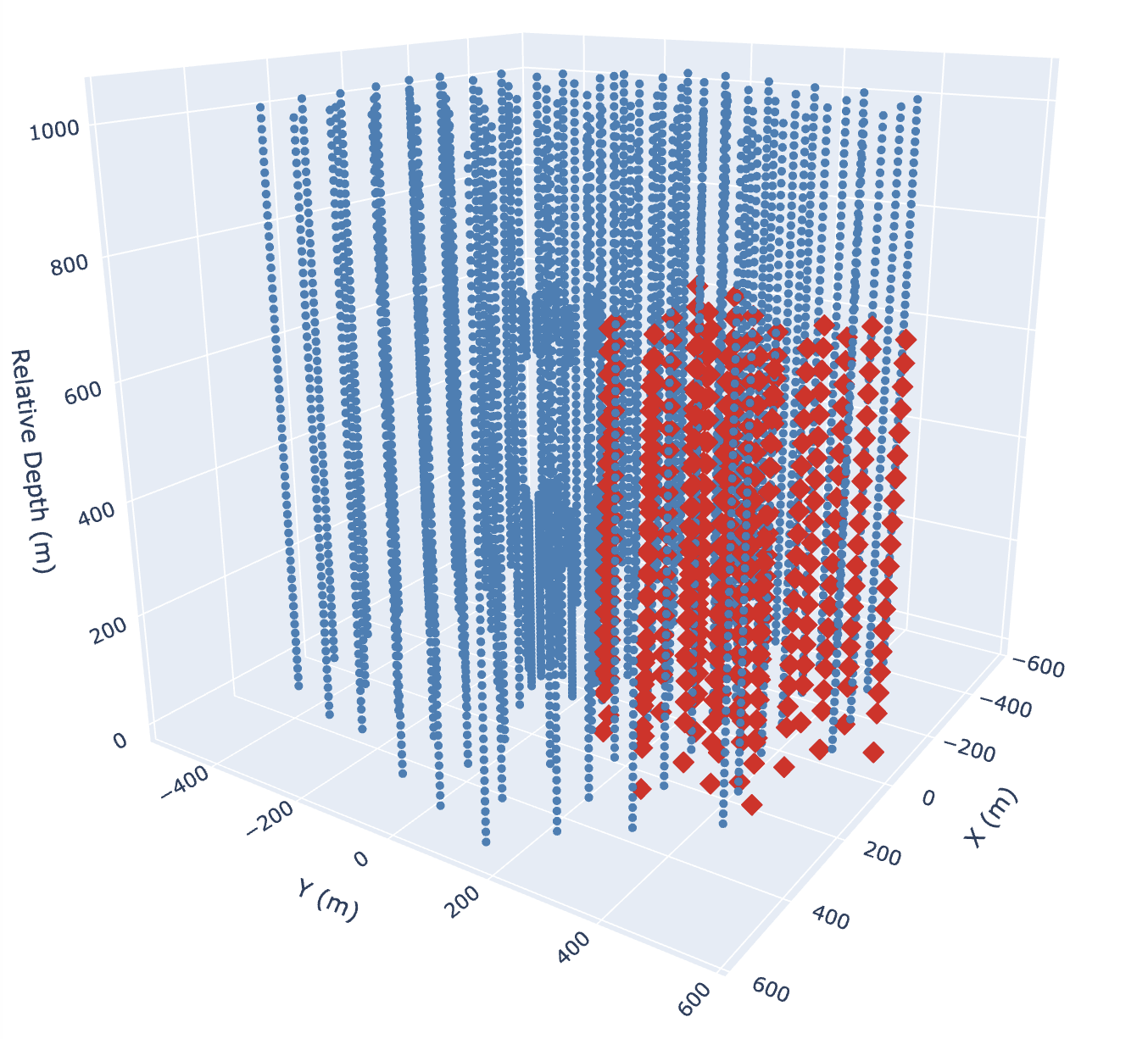}
  \caption{
    Comparison of IceCube and KM3NeT/ARCA21 detector geometries in their local coordinates. 
    The left panel shows a top-down ($x$--$y$) projection, while the right panel shows a 3D perspective view, highlighting the difference in instrumented volume. 
    IceCube DOMs are shown in blue circles, and KM3NeT/ARCA21 DOMs are marked with red diamonds. 
    In the $x$--$y$ projection, the KM3NeT/ARCA21 detection units are not point-like because of the tilt of the detection lines in water.
  }
  \label{fig:detector_geometry}
\end{figure}

A comparison of the detector geometries of IC86 and ARCA21 is shown in \cref{fig:detector_geometry}, 
projected in the horizontal (x--y) plane (left) and in three dimensions (right). 
The ARCA21 strings appear tilted due to ocean dynamics and detector design; each DU is anchored to the sea floor at its base, 
while the remainder of the cable is held upright by a buoy at the top, allowing them to sway with sea currents~\cite{KM3Net:2016zxf}. The relative positions of KM3NeT DOMs are precisely calibrated using a dedicated acoustic positioning system, 
which provides real-time measurements of their locations with sub-meter accuracy~\cite{KM3NeT:2021aqw}. 
This dynamic geometry needs to be accounted for in the reconstruction algorithms to obtain accurate event reconstruction and directional resolution.

The KM3NeT and IceCube detectors face fundamentally different operational conditions due to the contrasting properties of their detection media: deep seawater and Antarctic ice, respectively. A key logistical advantage of the KM3NeT design is the ability to deploy new DUs when necessary due to existing unit failures. In contrast, drilling and deploying new strings at the South Pole presents more logistical hurdles, and requires proper planning. Both environments pose distinct challenges for detector calibration and background mitigation. In the deep ocean, KM3NeT must contend with optical noise arising from bioluminescent marine organisms and the radioactive decay of potassium-40 ($^{40}$K) in seawater~\cite{KM3NeT_TDR_2008}. These effects introduce a persistent background photon rate, contribute to PMT aging, and must be accounted for in analyses. IceCube, on the other hand, benefits from a much lower optical noise environment but faces systematic uncertainties in the modeling of optical properties of the glacial ice, which are relevant for photon propagation~\cite{IceCube:2013llx,IceCube:2024qxf}. The Antarctic ice sheet has accumulated over hundreds of thousands of years and contains stratified layers that record geological and climatic history, including dust and volcanic ash~\cite{IceCube:2023qua}. These layers lead to depth-dependent optical properties that affect light propagation. To characterize the optical properties of the ice and calibrate the detector response, IceCube employs on-board LED flashers and laser systems distributed throughout the array. In contrast, seawater is more homogeneous in terms  of optical characteristics. However, as mentioned earlier the KM3NeT geometry is dynamically affected by sea currents, and requires continuous positional calibration via an acoustic tracking system.

Despite these contrasting conditions, both detectors offer complementary strengths in high-energy neutrino astronomy, with KM3NeT providing access to the Northern sky and IceCube to the Southern sky at neutrino energies beyond 10~PeV.

\subsection{Atmospheric, Astrophysical, and Cosmogenic Diffuse Neutrino Flux}

At IceCube, the typical data trigger rate is $\sim \SI{3}{\kilo \Hz}$~\cite{Aartsen:2016nxy}, dominated by atmospheric muons produced in cosmic-ray interactions with particles in the upper atmosphere~\cite{Gaisser2016_CRPhysics}. To reach the IceCube detector, which is embedded deep in the Antarctic ice, vertical muons must have a minimum energy of $\sim \SI{273}{\giga \eV}$~\cite{IceCube:2016umi} to overcome energy losses primarily due to ionization. This energy threshold increases for more inclined trajectories due to the greater slant depth through the ice.

The same cosmic-ray air showers that produce muons also produce neutrinos. These atmospheric neutrinos predominantly arise from the decay of charged pions and kaons~\cite{Gaisser2016_CRPhysics}. The primary decay channels are:
\begin{align*}
\pi^\pm &\rightarrow \mu^\pm + \nu_\mu \\
K^\pm &\rightarrow \mu^\pm + \nu_\mu \\
K^\pm &\rightarrow \pi^0 + e^\pm + \nu_e \\
K^0_L &\rightarrow \pi^\pm + e^\mp~(\mu^\mp) + \nu_e~(\nu_\mu) \\
\mu^\pm &\rightarrow e^\pm + \nu_e + \nu_\mu,
\end{align*}
where for brevity we have omitted the distinction between neutrino and antineutrino, which can be easily inferred. These channels contribute to the \emph{conventional atmospheric neutrino flux}~\cite{Gaisser2016_CRPhysics}, characterized by a steeply falling energy spectrum (\(\propto E^{-3.7}\)) and a dominance of muon neutrinos. At higher energies, most muons reach the ground where they lose energy catastrophically before decaying, further suppressing the \(\nu_e\) contribution. As a rough guideline, the conventional atmospheric \(\nu_e\) flux is \(\sim 10\) times lower than that of \(\nu_\mu\) at TeV energies~\cite{IceCube:2016umi}.

In addition to the conventional flux, there is a predicted component known as the \emph{prompt atmospheric neutrino flux}. Prompt neutrinos originate from the decay of short-lived heavy-flavoured mesons, primarily charm hadrons (e.g.~\(D^\pm\), \(D^0\), \(\Lambda_c^\pm\)), produced in cosmic-ray interactions~\cite{Enberg:2008te}. Due to their short lifetimes (\(\tau \sim 10^{-12}\)~s), charm hadrons decay promptly without undergoing further interactions in the atmosphere. As a result, the prompt neutrino spectrum is expected to follow the shape of the primary cosmic-ray spectrum more closely (\(\propto E^{-2.7}\)). Unlike the conventional flux, the prompt flux includes \(\nu_\mu\) and \(\nu_e\) in roughly equal proportions and is expected to become the dominant atmospheric component above around \SI{100}{\tera \eV}~\cite{Enberg:2008te}.

\begin{figure}[htbp]
  \centering
  \includegraphics[width=0.8
  \textwidth]{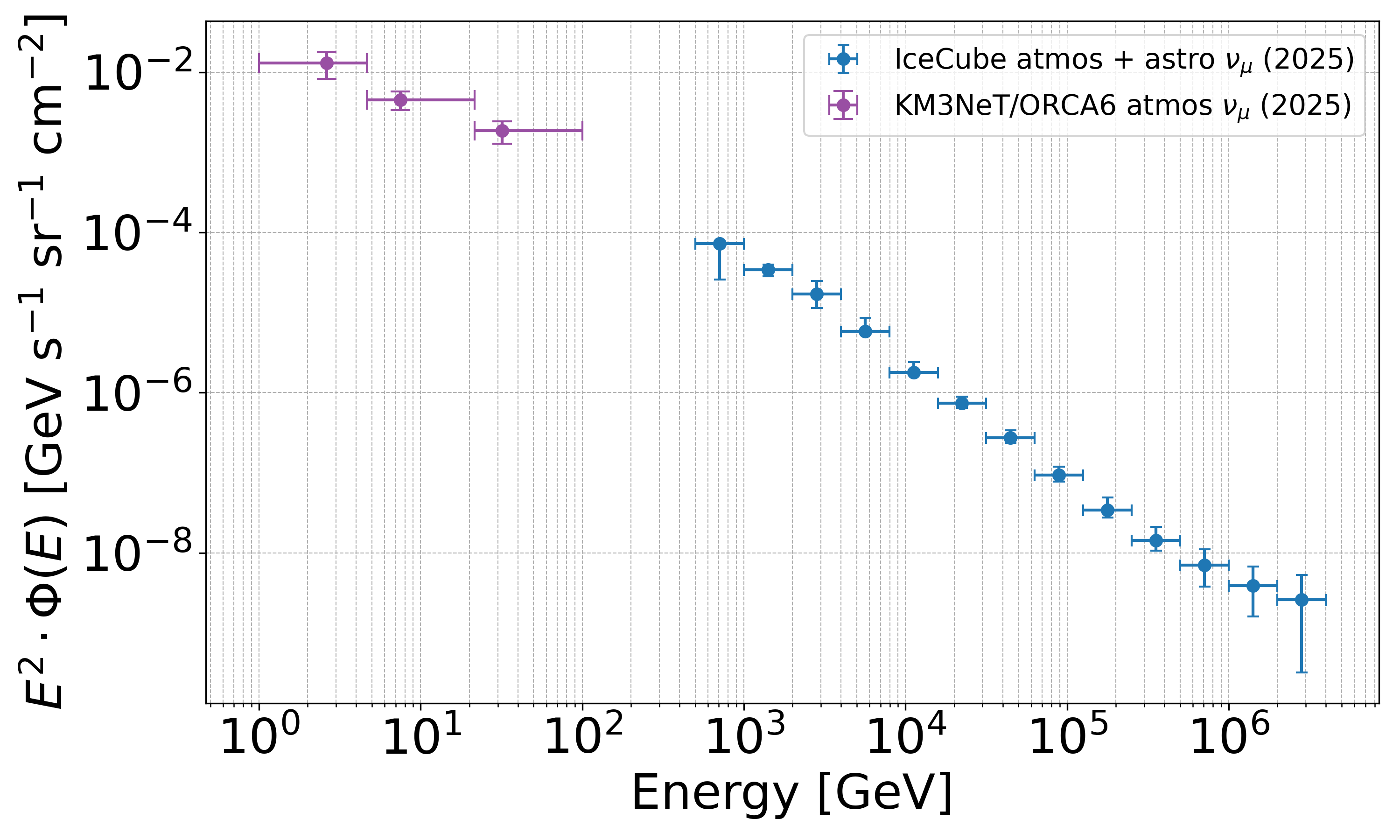}
    \caption{
    Comparison of atmospheric muon neutrino energy spectra measured by IceCube (blue) and KM3NeT/ORCA6 (purple). 
    The IceCube spectrum~\cite{VanRootselaar2025_IceCubeNuMu} is derived from through-going $\nu_\mu$ events collected over eleven years, covering an energy range from 500~GeV to 4~PeV. 
    It includes contributions from the conventional atmospheric component, the prompt component, and astrophysical component at the highest energies. 
    The ORCA6 spectrum~\cite{KM3NeT:2025pne} is based on a \SI{433}{\kilo ton \cdot years} exposure with the first six deployed detection units and covering an energy range from few~GeV to several tens of~GeV.
    }
    
  \label{fig:overview_diffuse_atmos}
\end{figure}
\Cref{fig:overview_diffuse_atmos} shows measurements of muon neutrino flux obtained by IceCube (IC86)~\cite{VanRootselaar2025_IceCubeNuMu} and of the atmospheric muon neutrino flux by KM3NeT/ORCA6~\cite{KM3NeT:2025pne}. The IceCube results are based on through-going muon neutrino events from the Northern Hemisphere and include contributions from the conventional atmospheric flux, the prompt component, and an additional astrophysical contribution at the highest energies. The ORCA6 atmospheric muon neutrino flux measurement is based on \SI{433}{\kilo ton \cdot years} of exposure collected with the first six deployed DUs. The analysis uses machine learning classifiers to select a high-purity up-going muon neutrino sample with atmospheric muon contamination suppressed to below 1\%. The ORCA6 measurement covers the energy range 1–100\,GeV and is primarily sensitive to the conventional atmospheric neutrino component. Within uncertainties, the measured muon neutrino fluxes from both IceCube and KM3NeT/ORCA6 appear to follow a consistent power-law trend where atmospheric neutrinos are expected to dominate. At energies above \SI{100}{\tera \eV}, the transition in slope is in-line with the existence of a diffuse flux of \emph{astrophysical neutrinos}. Other IceCube measurements of the diffuse astrophysical neutrino flux can be found in Ref.~\cite{IceCube:2020acn,IceCube:2020wum,IceCube:2021xar,IceCube:2025tgp,Lyu:2024jsm}, with energies extending beyond a PeV. These neutrinos are thought to originate from cosmic accelerators such as active galactic nuclei (AGNs), gamma-ray bursts (GRBs), and starburst galaxies. 

To date, IceCube has reported two notable source associations. A spatial clustering of neutrino events consistent with the Seyfert II galaxy NGC~1068 was observed with a post-trial significance of \(\sim 4\sigma\)~\cite{IceCube:2022der}, suggesting steady neutrino emission from this active galaxy. In the context of real-time multi-messenger follow-up observations, a high-energy neutrino alert was issued in coincidence with a multi-wavelength flare from the blazar TXS~0506+056~\cite{IceCube:2018dnn}. This event, detected across gamma-ray, optical, and radio bands, showed spatial and temporal alignment with the neutrino and yielded a combined significance of \(\sim 3.5\sigma\). While these results provide evidence for individual source associations, a statistically significant correlation with an entire population of sources—such as blazars, GRBs, or starburst galaxies—has yet to be established, and the origin of the majority of the astrophysical neutrino flux remains unknown.

\begin{figure}[htbp]
  \centering
  \includegraphics[width=0.9\textwidth]{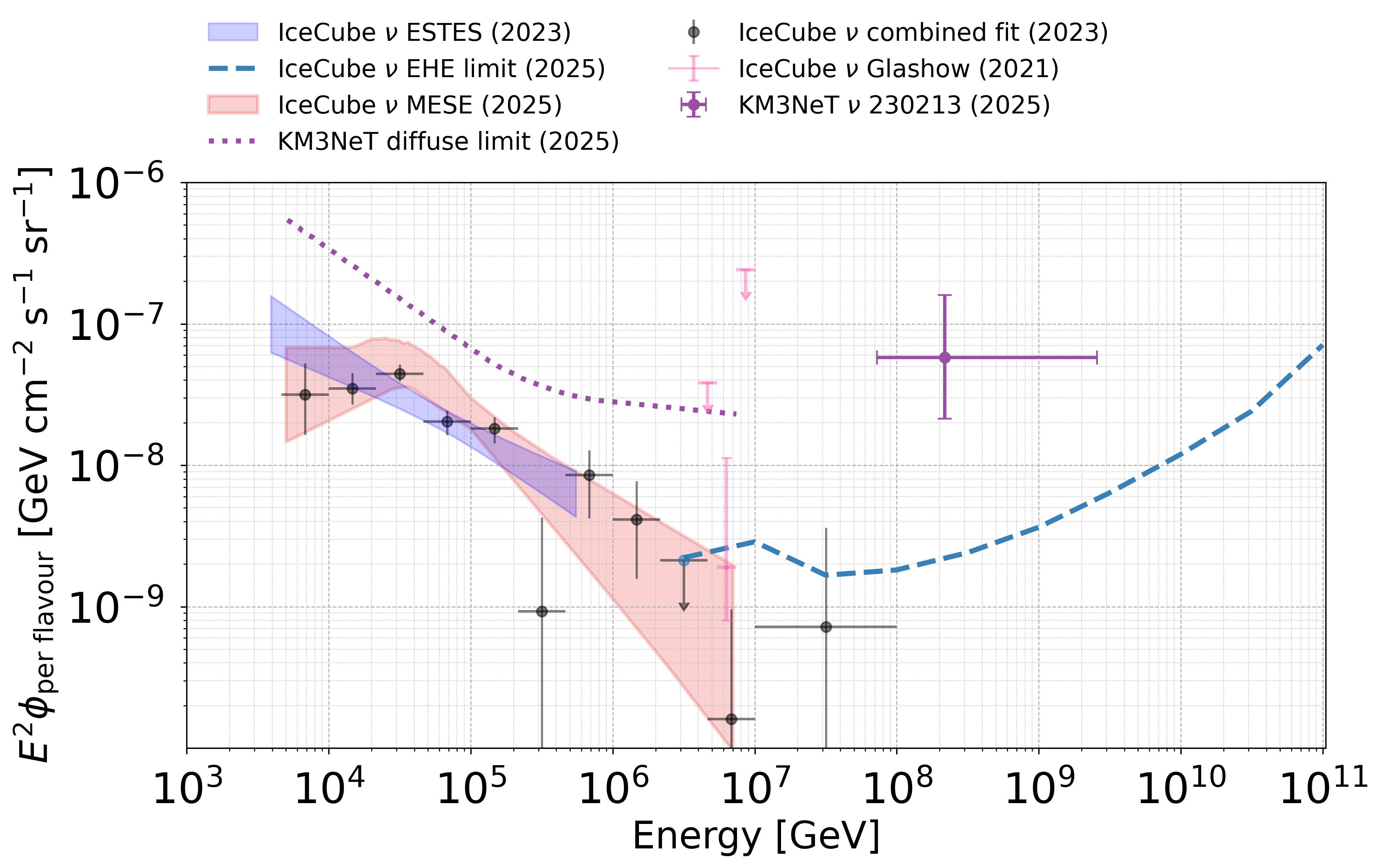}
    \caption{
    Overview of recent diffuse astrophysical neutrino measurements from IceCube and KM3NeT. 
    The blue shaded band shows the $1\sigma$ uncertainty region from the IceCube ESTES starting-track analysis~\cite{IceCube:2024fxo}, which employs ten years of data and prefers a single power-law spectrum. 
    The red shaded band shows the $1\sigma$ uncertainty from the IceCube MESE analysis~\cite{IceCube:2025tgp}, selecting both starting showers and tracks and favouring a broken power law. 
    Black points show the IceCube combined fit~\cite{IceCube:2025tgp}, which includes through-going tracks and starting showers and suggests spectral curvature. 
    The pink markers represent constraints from the Glashow resonance candidate event~\cite{IceCube:2021rpz}. 
    The dashed blue line is the IceCube EHE differential limit~\cite{IceCube:2025ezc}, 
    where the bump near $10^7$~GeV reflects the presence of three candidate events in that energy range, 
    whose characteristics are more compatible with an astrophysical-like spectrum than with a cosmogenic one.
    For KM3NeT, the dotted purple line shows the diffuse flux upper limit from ARCA~\cite{KM3NeT2025_DiffuseLimit}, and the purple error bar derived from the KM3-230213A event~\cite{KM3NeT:2025npi}.
    }

  \label{fig:overview_diffuse}
\end{figure}

The \emph{astrophysical flux} measured by IceCube using electron, muon, and tau neutrinos is shown in \cref{fig:overview_diffuse}. Analysis of a dataset of mostly tracks that start inside the detector suggests a continuous single power-law (SPL) spectrum extending from a few TeV up to several hundred TeV~\cite{IceCube:2024fxo}. However, analyses of shower events—primarily based on contained cascades—now reject the SPL model in favor a broken power-law (BPL) spectrum~\cite{IceCube:2025tgp}. The BPL model can be parameterized as:
\begin{align}
\Phi_{\nu+\bar{\nu}}(E_\nu) = \phi_{0, \text{broken}} \left( \frac{E_\nu}{E_{\text{break}}} \right)^{-\gamma_{\text{BPL}}}, \quad
\text{where} \quad
\gamma_{\text{BPL}} =
\begin{cases}
\gamma_1 & \text{for } E_\nu < E_{\text{break}} \\
\gamma_2 & \text{for } E_\nu > E_{\text{break}}
\end{cases}
\end{align}
with a normalization factor:
\begin{align}
\phi_{0,\text{broken}} = \phi_0 \times
\begin{cases}
\left( \frac{E_{\text{break}}}{100~\text{TeV}} \right)^{-\gamma_1} & \text{if } E_{\text{break}} > 100~\text{TeV} \\
\left( \frac{E_{\text{break}}}{100~\text{TeV}} \right)^{-\gamma_2} & \text{if } E_{\text{break}} \leq 100~\text{TeV}
\end{cases}
\end{align}
The best-fit parameters obtained from the IceCube diffuse combined fit are:
\begin{align*}
\phi_0 &= 1.77^{+0.19}_{-0.18} \times 10^{-18}~\text{GeV}^{-1}~\text{cm}^{-2}~\text{s}^{-1}~\text{sr}^{-1} \\
\gamma_1 &= 1.31^{+0.51}_{-1.30} \\
\gamma_2 &= 2.735^{+0.067}_{-0.075} \\
\log_{10}\left( \frac{E_{\text{break}}}{\text{GeV}} \right) &= 4.39^{+0.10}_{-0.10} \quad \Rightarrow \quad E_{\text{break}} \approx 24.5~\text{TeV}
\end{align*}
Updates to the Antarctic ice model~\cite{IceCube:2024qxf,IceCube:2023qua}, improved event reconstruction techniques~\cite{IceCube:2024csv}, and refined modeling of the atmospheric neutrino flux~\cite{Yanez:2023lsy} will be incorporated into the next iteration of diffuse astrophysical flux measurements, with the aim to resolve the discrepancies between showers and tracks~\cite{Yildizci:2025ult}.

In parallel, KM3NeT has reported an upper limit on the diffuse astrophysical neutrino flux based on early data from ARCA~\cite{Dornic2025_KM3NeTStatus}. This limit, shown as a dotted line in \cref{fig:overview_diffuse}, is consistent with the flux measured by IceCube~\cite{IceCube:2024fxo, IceCube:2025tgp}. A future measurement of the astrophysical flux by KM3NeT will provide an important independent cross-check via a different detection medium and offering complementary coverage of the neutrino sky.

At even higher energies (EeV scale), a fourth component—the \emph{cosmogenic neutrino flux}—is expected. These neutrinos are produced through interactions of UHECRs with CMB photons via the GZK mechanism, as illustrated in \cref{eq1}. Cosmogenic neutrinos provide a unique probe of UHECR composition, source distribution, and propagation over cosmological distances~\cite{vanVliet:2019nse}. IceCube upper limits on the flux in this energy regime are shown as the dashed blue line in \cref{fig:overview_diffuse}. KM3NeT/ARCA21 has currently detected one candidate event with a possible neutrino energy extending up to a EeV. The details and implications of this detection, as well as the detections of the highest-energy IceCube events, will be discussed in the next sections.

\section{Detecting EHE neutrinos at IceCube and KM3NeT}
\label{sec:detect}

\begin{figure}[htpb]
  \centering
  \includegraphics[width=0.49\textwidth]{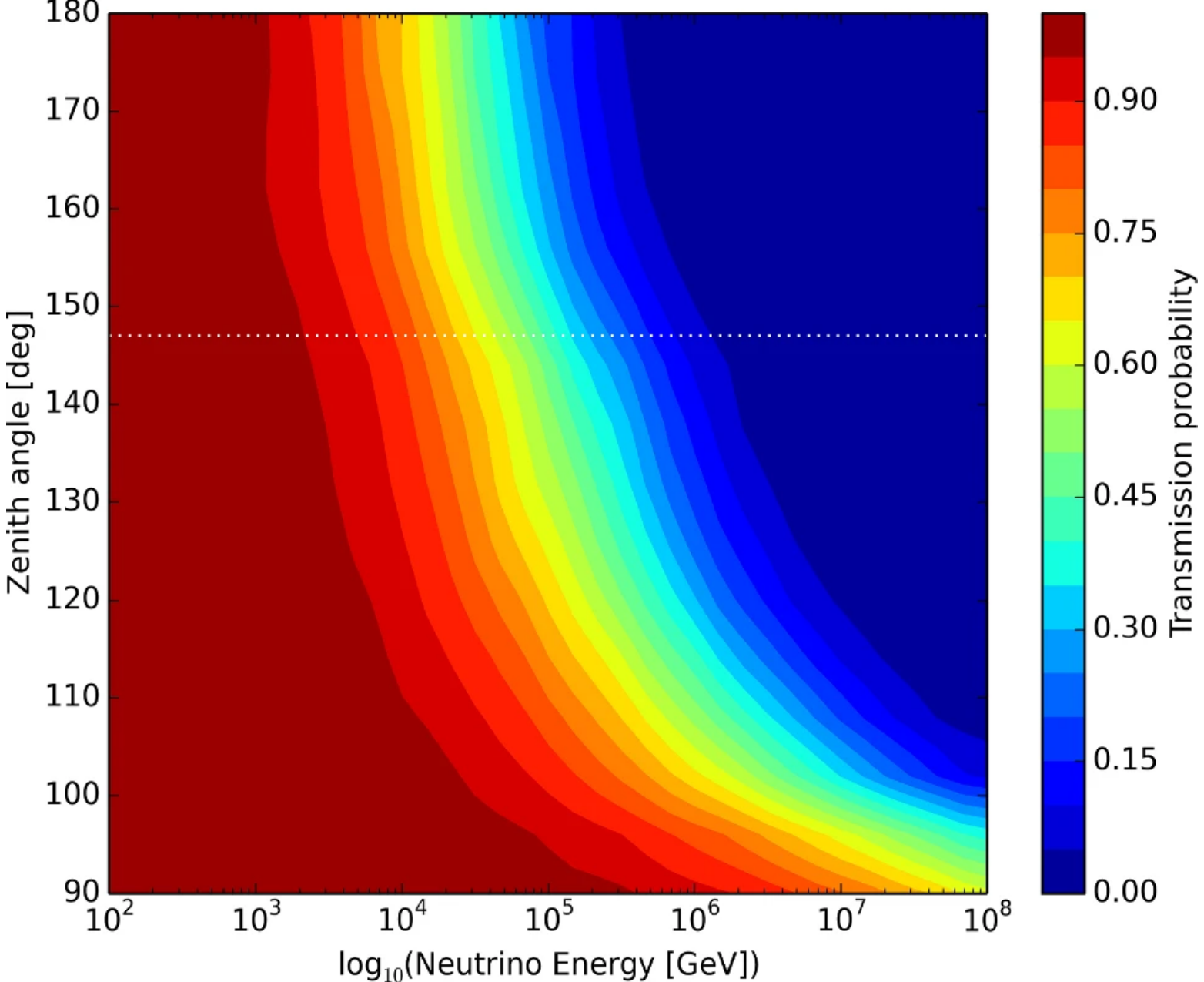}
  \hspace{1cm}
  \includegraphics[width=0.23\textwidth]{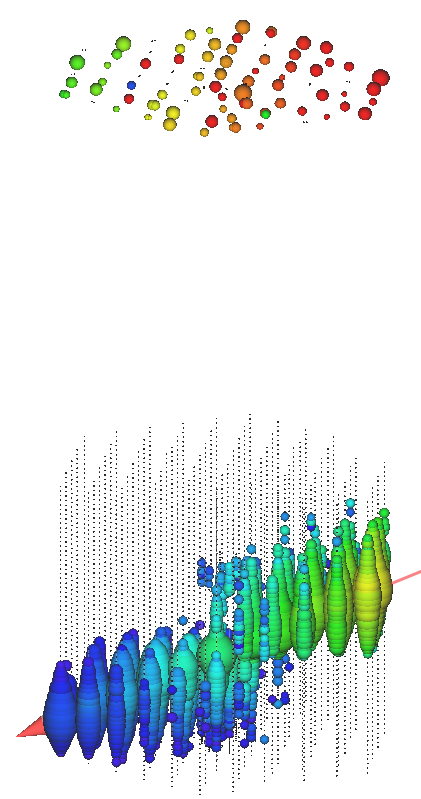}
  \caption{
    Left: Neutrino transmission probability through the Earth as a function of energy and zenith angle~\cite{IceCube:2017roe}. 
    A zenith angle of \ang{180} corresponds to a trajectory passing through the full diameter of the Earth. 
    Right: Event display of IC200728A, an atmospheric air shower with both an in-ice component (lower panel) and time-coincident signatures in the IceTop surface array~\cite{IceCube:2012nn} (upper panel). 
    Without the IceTop information, such an event would mimic a neutrino interaction in IceCube, but is in reality a background event vetoed by the surface detector.
    }
    
  \label{fig:survival_veto_combo}
\end{figure}

Extremely high-energy neutrinos, typically defined as those with energies exceeding $\SI{\sim 5}{\peta \eV}$, pose unique detection challenges. Due to the increasing neutrino-nucleon cross section at these energies, the Earth becomes effectively opaque to neutrinos traversing large distances~\cite{IceCube:2017roe}, as illustrated in the left panel of \cref{fig:survival_veto_combo}. 
As a result, EHE neutrinos are predominantly observed from near-horizontal or downward-going directions, eliminating the possibility of using the Earth as a natural shield to suppress the background from atmospheric muons.

Electron neutrino charged current (CC) interactions, as well as neutral current (NC) interactions of all flavours, typically produce shower-like events in the detector. These are more easily distinguished from atmospheric muons, which appear as long, track-like signatures. However, the effective area for detecting showers is significantly smaller than that for tracks, which often travel kilometers through the medium, allowing for detection of interactions that occur far outside the instrumented volume. Consequently, at EHEs muon and tau neutrinos have larger effective areas due to their penetrating secondaries~\cite{IceCube:2018fhm,IceCube:2025ezc}. This advantage, however, comes at the cost of more challenging background rejection, as the interaction vertex often lies outside the detector, making it more difficult to distinguish neutrino-induced tracks from atmospheric muons.

\subsection{Techniques for muon background rejection}

To first order, the deposited energy in the detector correlates with the total amount of Cherenkov light, which can be approximated by detector-specific observables. In IceCube, the PMT signals are digitized using both Analog Transient Waveform Digitizers (ATWD) and Fast Analog-to-Digital Converters (FADC), providing full waveform information~\cite{Aartsen:2016nxy}. This allows for precise charge integration and a more accurate estimate of the event energy. 

\begin{figure}[ht]
    \centering
    \includegraphics[width=0.47\textwidth]{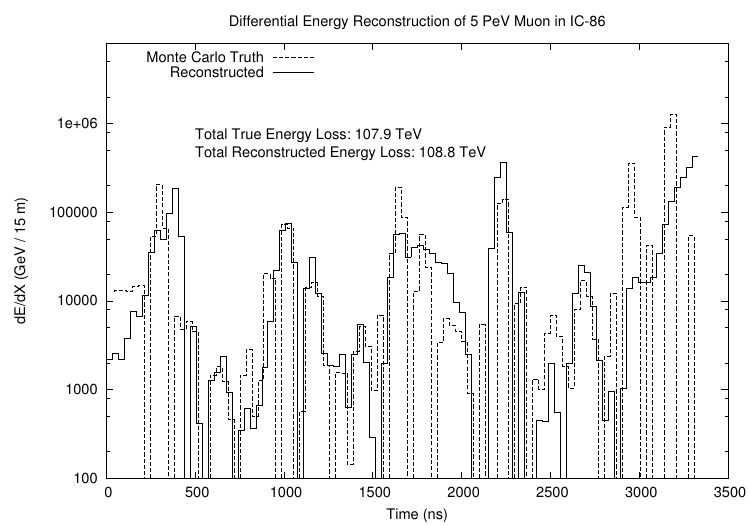}
    \includegraphics[width=0.47\textwidth]{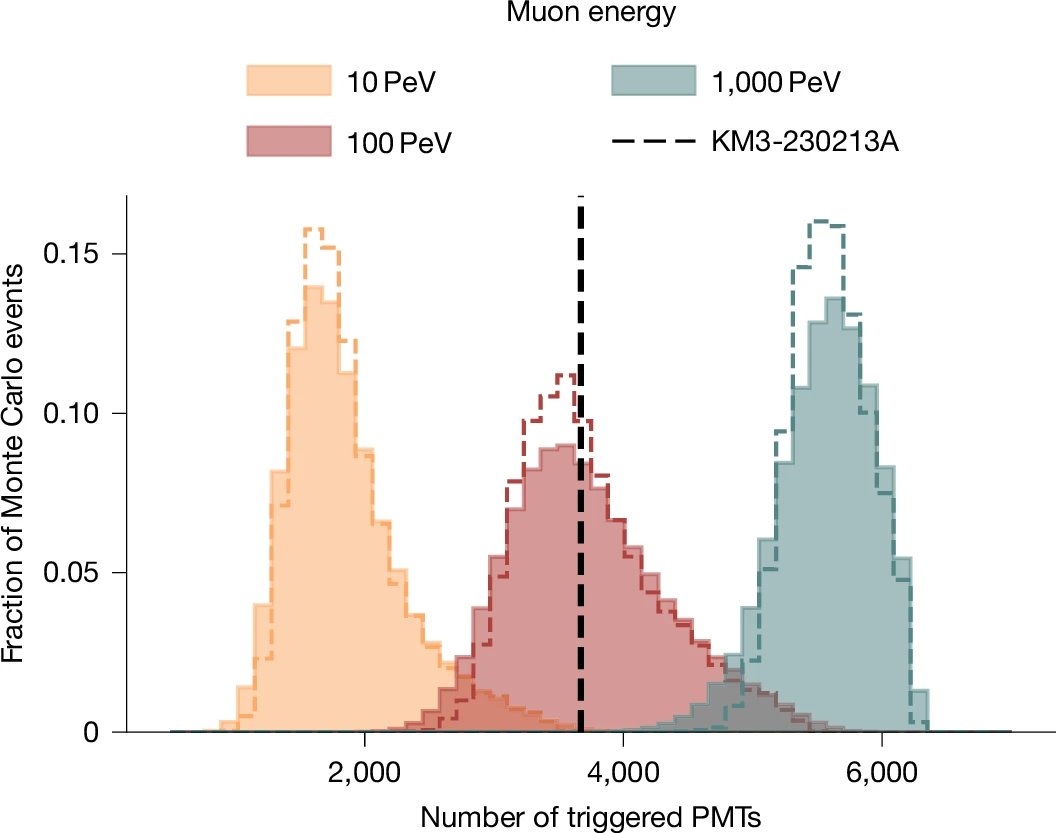}

    \caption{
    Comparison of energy reconstruction methods in IceCube and KM3NeT. 
    Left: In IceCube, the Millipede algorithm~\cite{IceCube:2013dkx} reconstructs the energy deposition profile of a muon track by segmenting it into short intervals and fitting the cascade energy at each segment to the observed photoelectron time distributions over all DOMs. 
    The resulting per-segment deposited energies can then be used to derive track-based energy estimators, such as $\dd E/\dd x$~\cite{IceCube:2012iea} or the total deposited energy. 
    Right: For KM3-230213A, the energy was reconstructed by correlating the total number of PMTs registering hits with MC simulations of different muon energies, accounting for detector geometry and optical water properties.
    }
\label{fig:energy_reco_comparison}
\end{figure}

On the other hand,  KM3NeT employs a time-over-threshold (ToT) readout system, in which only the time duration that a signal remains above a predefined voltage threshold is recorded~\cite{KM3NeT:2022pnv}. While this approach enables a compact and efficient readout for the 31 small PMTs housed in each DOM, it does not capture the full waveform or integrated charge. As a result, the number of triggered PMTs was used in~\cite{KM3NeT:2025npi} to estimate the energy of KM3-230213A (c.f.~\cref{fig:energy_reco_comparison}).

More sophisticated methods exist for energy estimation~\cite{IceCube:2013dkx}, particularly for track-like events in IceCube. One commonly used approach is to infer the muon energy from its differential energy loss (\( \dd E/\dd x \), c.f.~left panel of \cref{fig:energy_reco_comparison}) as it traverses the detector~\cite{IceCube:2012iea}. This quantity represents the rate at which energy is deposited per unit path length. At high energies, muon energy loss is dominated by stochastic processes such as bremsstrahlung, pair production, and photonuclear interactions, which scale approximately linearly with energy. Consequently, the average Cherenkov light yield per unit length becomes an effective proxy for the muon energy. 

In addition to energy reconstruction, the stochastic nature of high-energy muon energy loss provides a means to discriminate neutrino-induced single muons from background muons produced in cosmic-ray air showers~\cite{IceCube:2025ezc}. Muons reaching the detector with energies above \(\sim10\,\mathrm{PeV}\) are typically part of muon bundles rather than isolated tracks. These bundles result from multiple muons propagating in parallel, each depositing energy independently. As a consequence, the total energy loss per unit length becomes a sum of many smaller, uncorrelated losses—effectively smoothing out the overall \(\dd E/\dd x\) profile. In contrast, a single high-energy muon undergoes rare but large stochastic losses, leading to high fluctuations in deposited energy. This difference in \( \dd E/\dd x \) characteristics can therefore be exploited to suppress the atmospheric muon background in EHE neutrino searches.

Traditionally, the selection of EHE neutrino events has relied on identifying bright events across all neutrino flavours, using zenith-angle-dependent charge thresholds to maintain a low background rate~\cite{IceCube:2018fhm}. Building on the stochastic energy loss characteristics discussed above, the selection can be further refined by targeting events with pronounced localized energy deposits along their tracks. This allows for a relaxation of the total charge requirement in regions with high stochasticity, thereby improving sensitivity—particularly to down-going events from the Southern sky, where EHE neutrinos are more likely to survive propagations due to Earth absorption~\cite{IceCube:2025ezc}.

\begin{figure}[htbp]
  \centering
  \includegraphics[width=0.95\textwidth]{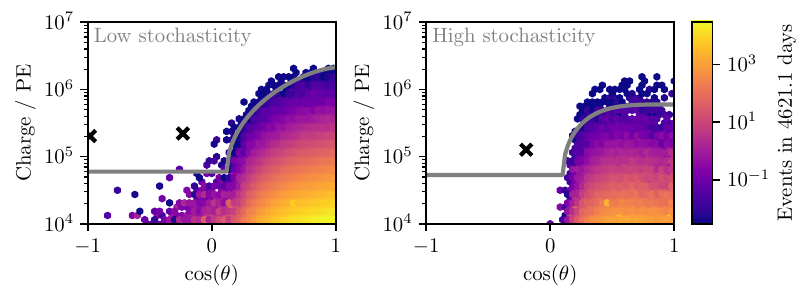}

    \caption{
    Background expectations for the EHE neutrino selection in IceCube, based on 4621.1 days of data~\cite{IceCube:2025ezc}. 
    Left: Low-stochasticity events, which are more consistent with cosmic-ray air-shower muons than with neutrino-induced stochastic losses (i.e., single muons from deep inelastic scattering). 
    Right: High-stochasticity events, where the selection threshold is less strict in order to retain high-charge events from the Southern sky. 
    The three data points (crosses) correspond to the candidate events discussed in \cref{sec:icecube_ehe}. 
    Note that the zenith angles shown are the reconstruction proxies used in this specific analysis; more sophisticated reconstructions were performed for the three events, indicating that both IC190331A and the Glashow resonance candidate originated from the Southern sky.
    }

  \label{fig:icecube_ehe_bkg}
\end{figure}

\begin{figure}[htbp]
  \centering
  \begin{minipage}{0.9\textwidth} 
    \centering
    \begin{subfigure}[t]{0.35\textwidth}
      \centering
      \includegraphics[width=\textwidth]{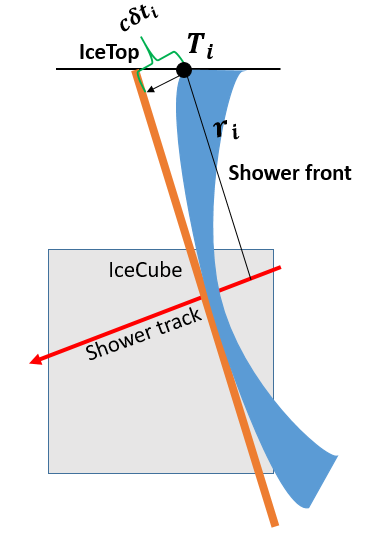}
    \end{subfigure}
    \hfill
    \begin{subfigure}[t]{0.55\textwidth}
      \centering
      \includegraphics[width=\textwidth]{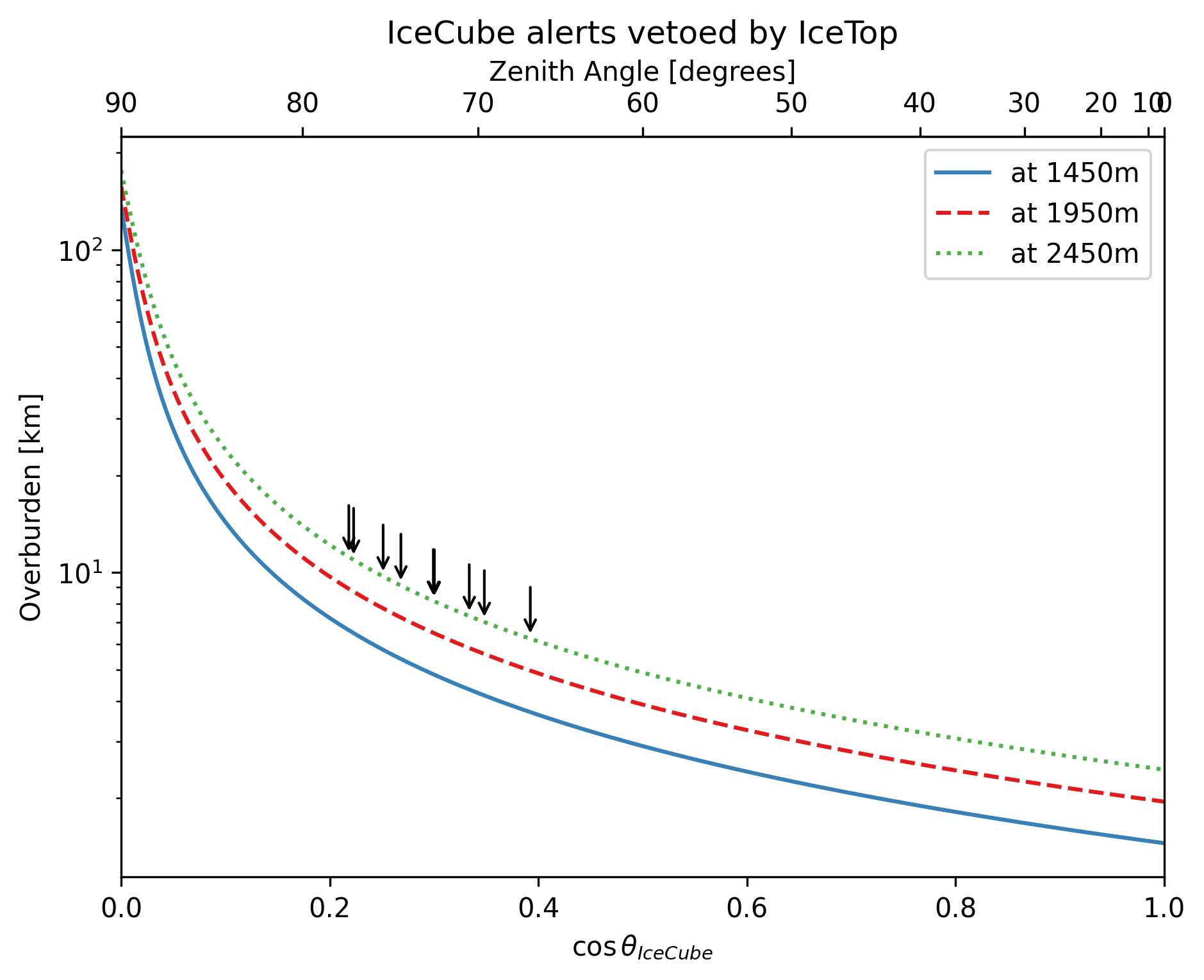}
    \end{subfigure}
  \end{minipage}
    \caption{
    Using the IceTop surface array as a veto for cosmic-ray background. 
    Left: Geometry of IceTop and IceCube, illustrating how IceTop can reject down-going background events. 
    The time difference between signals in the in-ice array and at the surface must be consistent with originating from the same extensive air shower. 
    Right: Overburden (slant depth) as a function of zenith angle at IceCube for three different detector depths, as indicated by the three curves. 
    Black arrows mark eight IceCube real-time alerts that were vetoed by IceTop. 
    Left panel courtesy of N.~M.~Binte~Amin.
    }

  \label{fig:icetop_veto_bc}
\end{figure}

An additional method for rejecting cosmic-ray background in IceCube is the use of the IceTop surface array, consisting of ice-Cherenkov tanks located at the South Pole, 1450~m above the top layer of the in-ice DOMs~\cite{IceCube:2012nn}. IceTop is sensitive to the electromagnetic and hadronic components of extensive air showers, making it an effective veto for down-going events. Background rejection is achieved by identifying coincident signals in IceTop and the in-ice array and requiring that surface hits are causally connected and geometrically consistent with the reconstructed trajectory of the in-ice event (left panel, \cref{fig:icetop_veto_bc})~\cite{Lyu:2024jsm}. This hybrid detection approach is also employed in IceCube’s real-time alert program~\cite{IceCube:2016cqr,Blaufuss:2019fgv}, where it significantly reduces background contamination in high-energy muon neutrino alerts. 

The right panel of \cref{fig:icetop_veto_bc} shows the atmospheric muon overburden before reaching the IceCube detector as a function of zenith angle, with \(0^{\circ}\) corresponding to vertically down-going trajectories from the atmosphere directly into the in-ice array. While the surface veto is most efficient for zenith angles up to \(30^{\circ}\), the figure illustrates its significant impact even for more inclined background events. Arrows indicate real-time alerts with reconstructed neutrino energies well above 100~TeV that were rejected by the surface veto. The largest zenith angle for which a vetoed event has been recorded to date is \(80^{\circ}\), only \(10^{\circ}\) above the horizon. An example event view for such a vetoed event, with a reconstructed zenith angle of \(65^{\circ}\), is shown in the right panel of \cref{fig:survival_veto_combo}.

Similarly, KM3NeT has developed a bright-track selection~\cite{KM3NeT:2025npi} requiring more than 1500 triggered PMTs, a reconstructed track length exceeding 250\,m, and a track-reconstruction log-likelihood ratio greater than 500. Such a selection is designed to efficiently suppress cosmic-ray--induced backgrounds while retaining high-energy neutrino candidates.

\begin{figure}[htbp]
  \centering
  \includegraphics[width=0.48\textwidth]{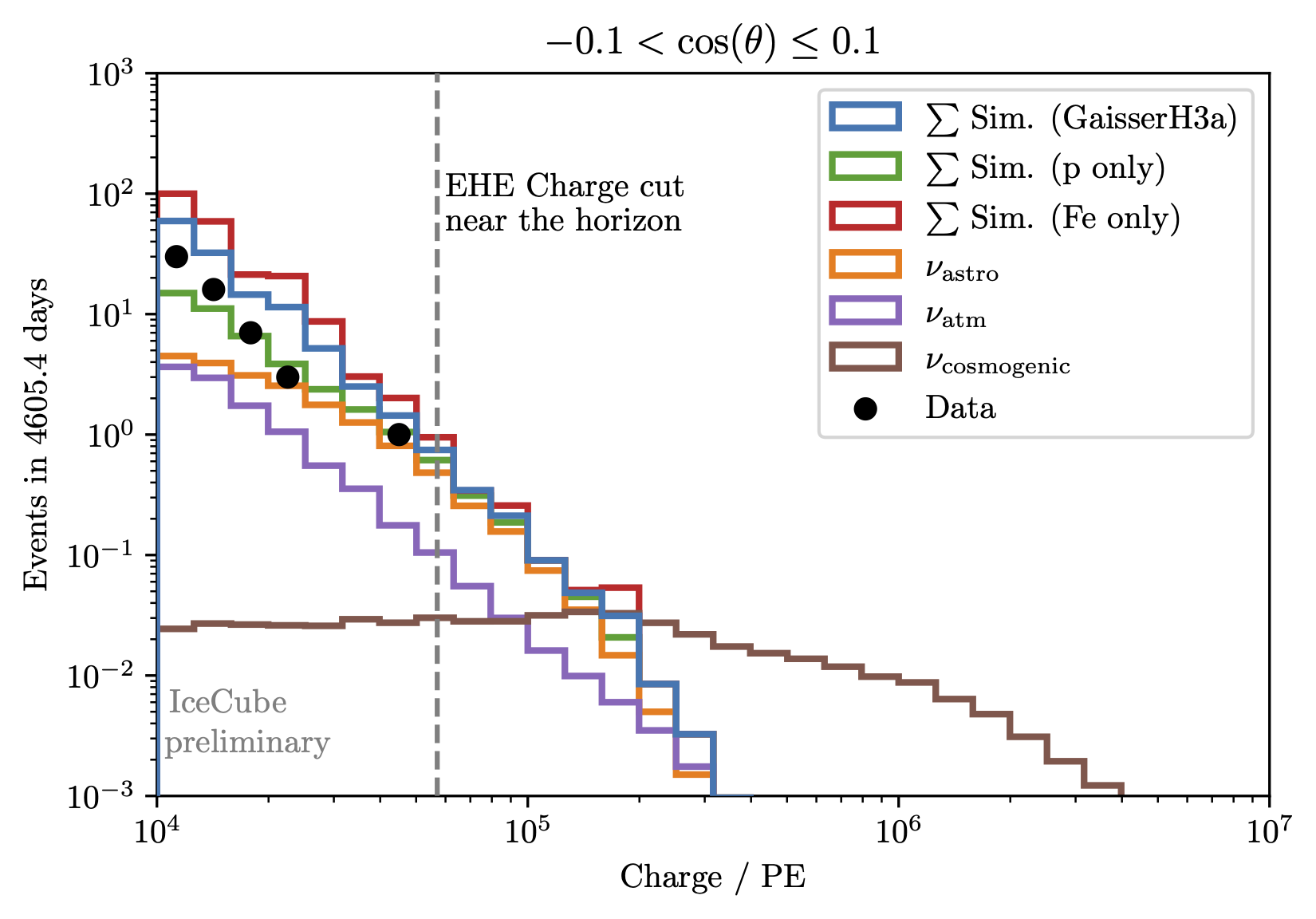}
  \includegraphics[width=0.44\textwidth]{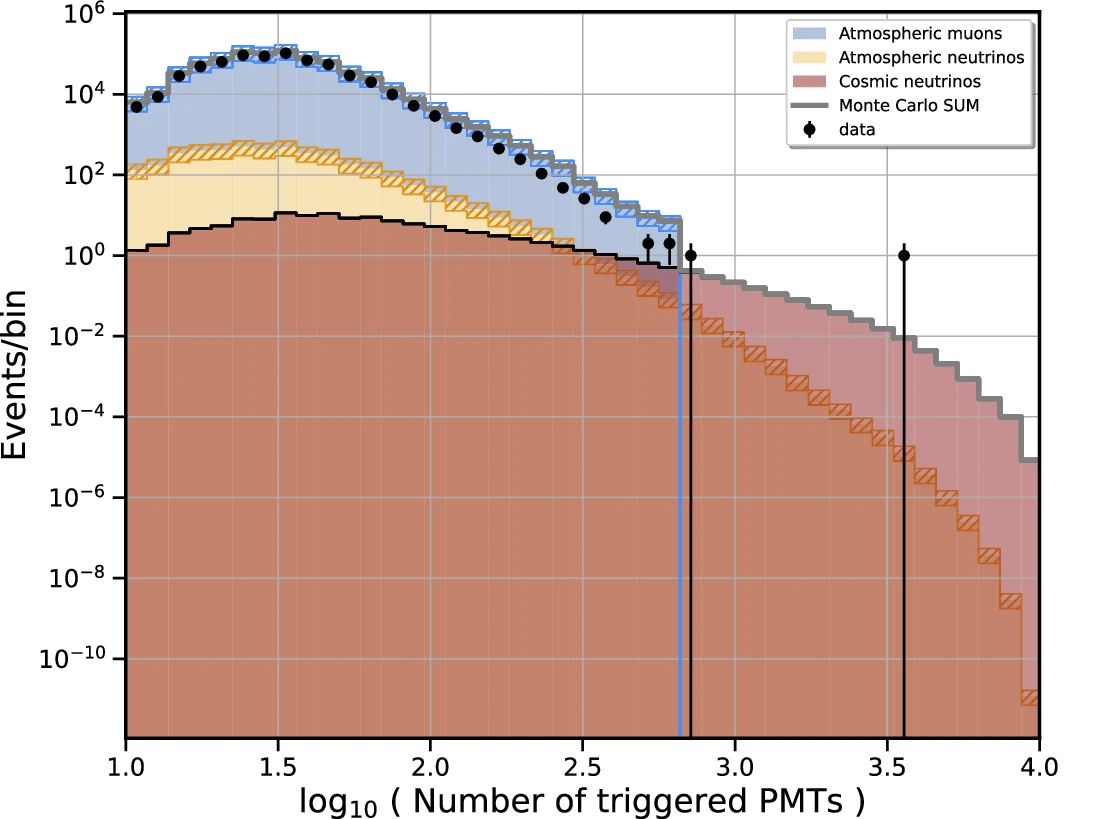}
    \caption{
    Data/MC comparisons for IceCube (left) and KM3NeT (right). 
    Left: Sub-threshold EHE event selection in IceCube~\cite{Meier:2025nql}. 
    The dashed line indicates where the EHE total charge cut lies near the horizon. 
    The distribution shown is for events in the horizontal zenith bin ($-0.1 < \cos\theta < 0.1$), chosen to match the reconstructed zenith angle of KM3-230213A. 
    At lower charges, the distribution is dominated by cosmic-ray air-shower muons whose directions are misreconstructed. 
    The data lie between the proton-only and iron-only composition extremes in the simulation. 
    No events survived the final EHE selection in this zenith range. 
    Right: Full-sky data/MC comparison of the number of triggered PMTs in the ARCA21 analysis, with KM3-230213A appearing as the outlier point in the data~\cite{KM3NeT:2025npi}. 
    The KM3NeT MC sample shown here is pre-fit.
    }

  \label{fig:icecube_km3_bkg}
\end{figure}

In both collaborations, diffuse neutrino analyses employ a forward-folding approach, fitting the observed data to MC predictions that include detector response and atmospheric-flux systematics. Achieving reliable data--MC agreement is therefore a prerequisite for unbiased flux measurements. For IceCube, an example is shown in the left panel of \cref{fig:icecube_km3_bkg}: the dashed line marks the location of the EHE charge cut near the horizon. The figure corresponds to a near-horizontal zenith bin with $|\cos\theta|<0.1$, where the low-charge region is dominated by atmospheric muons and the high-charge region by cosmogenic neutrinos. Below the selection threshold, the data agree well with the expected atmospheric-muon background, modeled for both proton and iron primary compositions using the SIBYLL~2.3 hadronic interaction model~\cite{Riehn:2019jet}.

A corresponding data-MC comparison for ARCA21 is shown in the right panel of \cref{fig:icecube_km3_bkg}. Here, a 40\% detector systematic due to water absorption has been applied, affecting the overall normalization of the histograms. This distribution is integrated over the whole sky, with cosmic neutrinos starting to dominate over cosmic-ray muons at around $N_{\mathrm{PMT}}\approx 500$. The sharp cutoff in the simulated atmospheric-muon distribution may indicate limited background statistics. Nevertheless, the final bright-track cut at $N_{\mathrm{PMT}}>1500$ lies well beyond the region where atmospheric muons are expected. At low $N_{\mathrm{PMT}}$ the data and MC are consistent, while above $\sim 100$~PMTs discrepancies appear; these could potentially be reduced by including nuisance parameters for the atmospheric-neutrino flux, the cosmic-ray mass composition, and hadronic-interaction models in the fit.

\begin{figure}[htbp]
  \centering
  \includegraphics[width=0.6\textwidth]{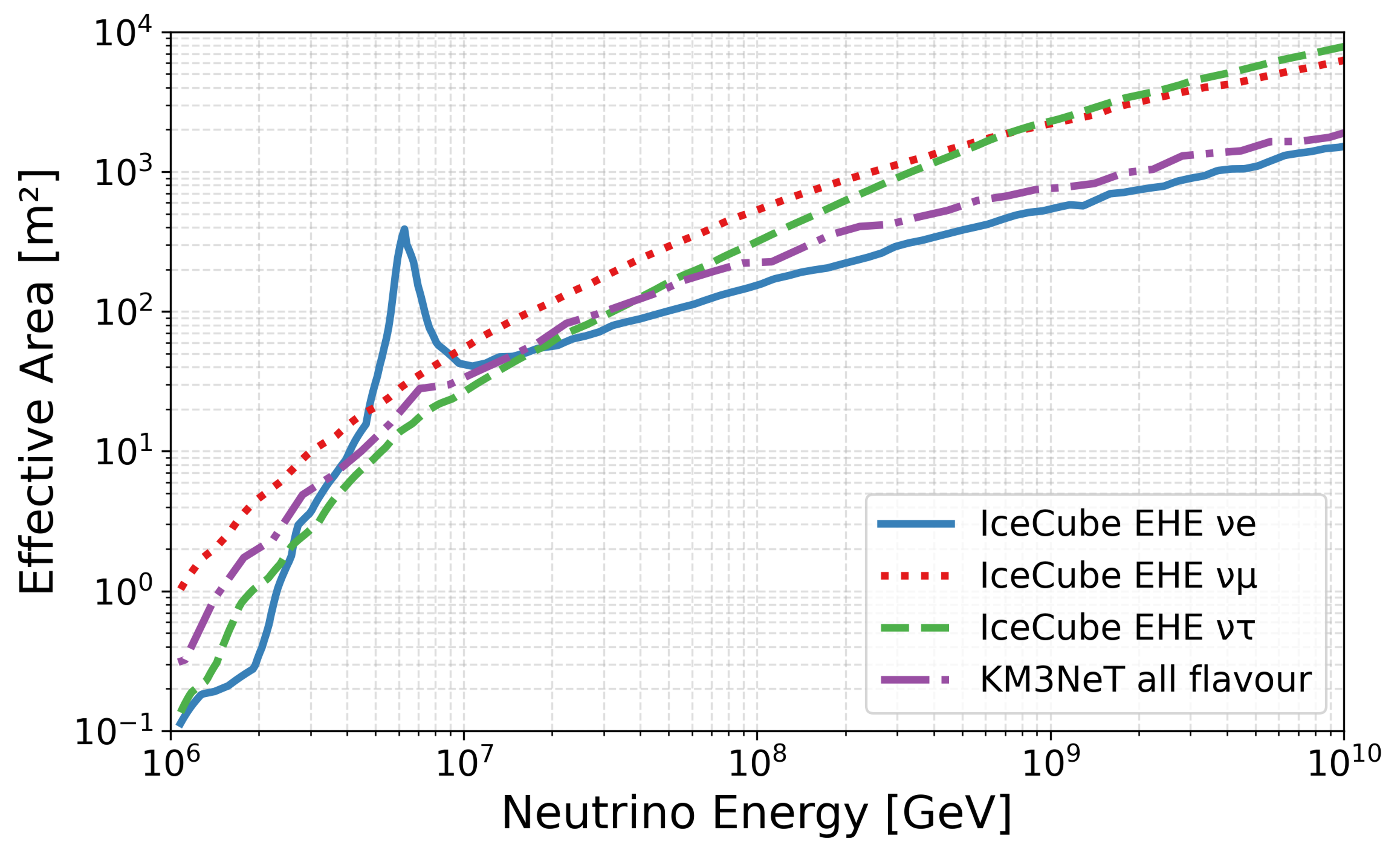}
    \caption{
    Comparison of effective areas for EHE neutrino selections in IceCube and KM3NeT. 
    The solid blue curve shows the $\nu_e$ and $\bar{\nu}_e$ effective area, with the visible enhancement from the Glashow resonance at 6.3~PeV. 
    The IceCube EHE selection~\cite{IceCube:2025ezc} is primarily sensitive to bright tracks but also shower events. 
    The red dotted curve corresponds to $\nu_\mu$, while the green dashed curve shows the $\nu_\tau$ effective area. 
    The purple dash-dotted curve shows the all-flavour effective area from KM3NeT's bright-track selection~\cite{KM3NeT:2025npi}.
    }
    
  \label{fig:effective_area}
\end{figure}
A comparison of the effective areas for the EHE selections in IceCube and KM3NeT is shown in \cref{fig:effective_area}. The IceCube selection is sensitive to all neutrino flavours. Shower-like events are dominated by interactions with vertices inside or just outside the instrumented volume, which also provides sensitivity to the Glashow resonance---the resonant interaction of an electron antineutrino with an electron producing an on-shell $W$ boson. At the highest energies, the effective areas for muon and tau neutrinos become comparable; for example, at $10^{18}$~eV the tau lepton decay length is about 50~km, increasing the effective volume for detecting stochastic tau events. For KM3NeT, the effective area shown here is integrated over all flavours. A flavour-dependent effective area was not available at the time of this review.

\subsection{Results of EHE neutrino observations from IceCube}
\label{sec:icecube_ehe}
\begin{figure}[htbp]
  \centering
  \begin{subfigure}[b]{0.32\textwidth}
    \centering
    \includegraphics[width=\textwidth]{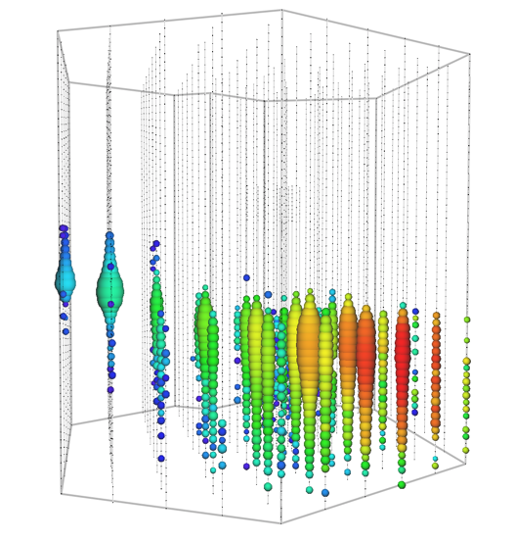}
    \caption{IC-20140611}
    \label{fig:evt1}
  \end{subfigure}
  \hfill
  \begin{subfigure}[b]{0.3\textwidth}
    \centering
    \includegraphics[width=\textwidth]{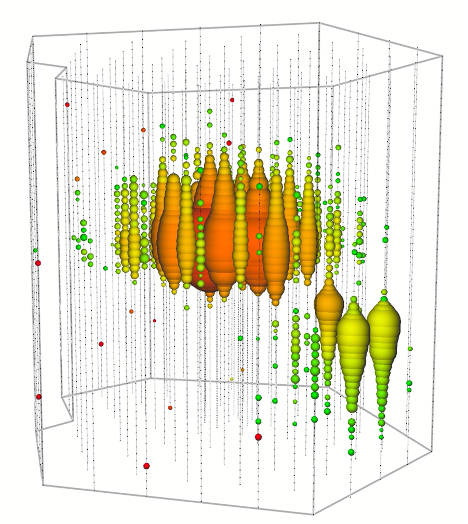}
    \caption{IC-20190331A}
    \label{fig:evt3}
  \end{subfigure}
  \hfill
  \begin{subfigure}[b]{0.36 \textwidth}
    \centering
    \includegraphics[width=\textwidth]{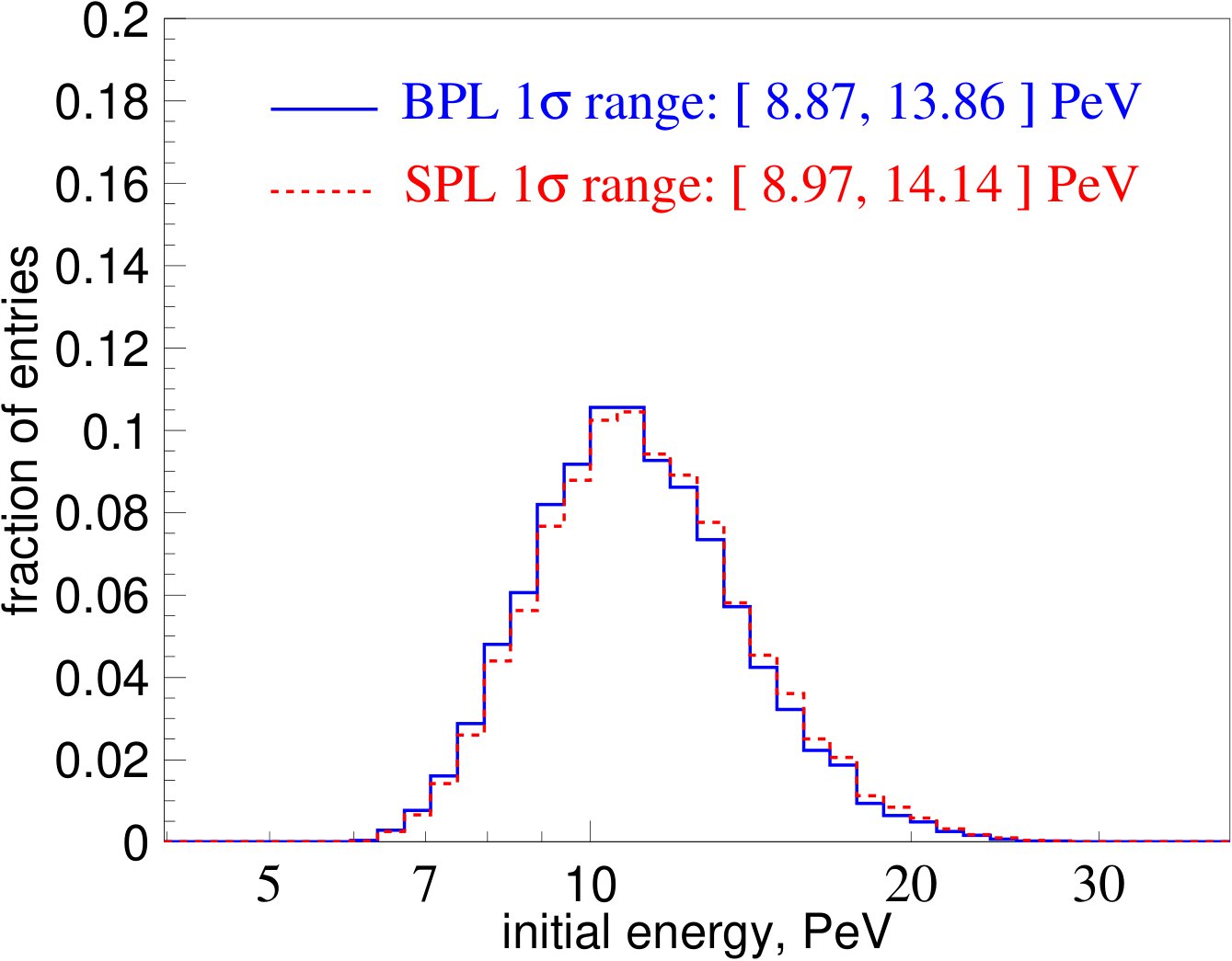}
    \caption{Inferred $E_\nu$ of IC-20190331A}
    \label{fig:evt3_energy}
  \end{subfigure}
  \caption{Two of the high energy events~\cite{IceCube:2016umi,IceCube:2025tgp} in IceCube with most probable neutrino energy above \SI{5}{\peta \eV}.}
  \label{fig:event_gallery}
\end{figure}

\begin{figure}[htbp]
  \centering
  \begin{minipage}{0.8\textwidth}
    \centering
    \begin{subfigure}[b]{0.35 \textwidth}
      \centering
      \includegraphics[width=\textwidth]{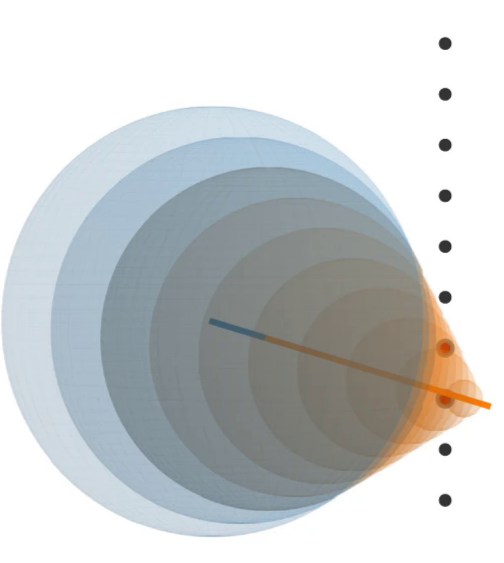}
      \caption{Early muon from hadronic shower travelling ahead of Cherenkov wavefront from the EM shower}
      \label{fig:rect1}
    \end{subfigure}
    \hfill
    \begin{subfigure}[b]{0.48\textwidth}
      \centering
      \includegraphics[width=\textwidth]{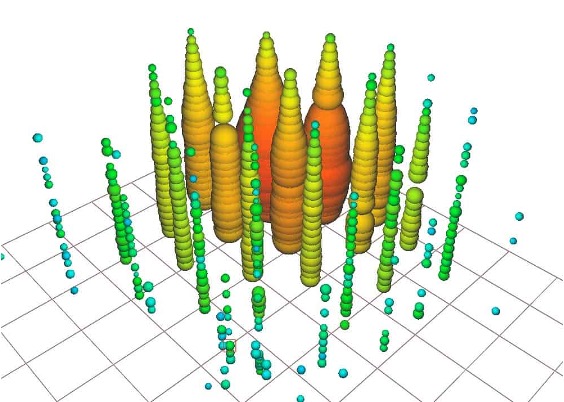}
    \caption{
    IC-20161208: Partially contained cascade event with an inferred neutrino energy of $\sim$6.3~PeV. 
    }

      \label{fig:evt2}
    \end{subfigure}
  \end{minipage}

  \vspace{0.8em} 
  \begin{subfigure}[b]{0.8\textwidth}
    \centering
    \includegraphics[width=\textwidth]{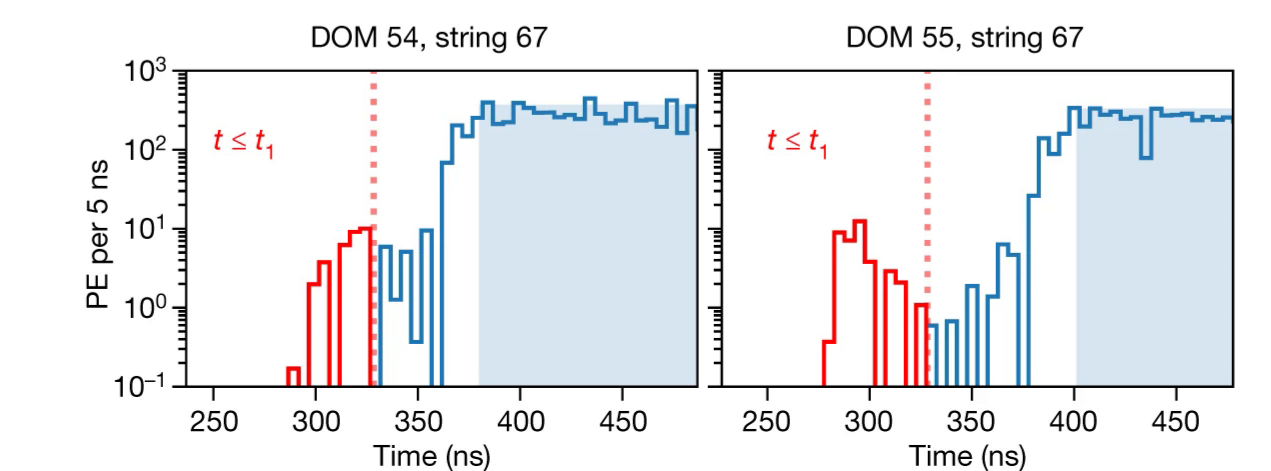}
    \caption{
    Example waveforms from two of the brightest DOMs recorded during the event. 
    Each panel shows the measured number of photoelectrons as a function of time. 
    The red histogram bins indicate hits that arrive too early to be explained by a purely electromagnetic hypothesis. 
    The large integrated charge reflects the brightness of the event, while the blue shaded regions mark portions of the waveform where the DOM readout was saturated.
    }
    \label{fig:waveform}
  \end{subfigure}

    \caption{
    The Glashow resonance candidate event~\cite{IceCube:2021rpz} and the accompanying muons produced in its hadronic cascade.
    }

  \label{fig:event_panel}
\end{figure}

A total of 12.4~years of IceCube data, from June 2010 to June 2023 and corresponding to 4605~days of livetime, were analyzed using the EHE event selection algorithm~\cite{IceCube:2025ezc}. Three events passed the final selection criteria and are shown in Fig.~\ref{fig:evt1}, Fig.~\ref{fig:evt2}, and Fig.~\ref{fig:evt3}. These are the highest-energy neutrino candidates detected by IceCube to date. Notably, all three events were also independently identified in dedicated analyses targeting distinct topologies: through-going tracks~\cite{Abbasi:2021qfz}, starting events~\cite{IceCube:2020wum}, and partially contained cascades~\cite{IceCube:2021xar}.

Though superior in directional resolution, with uncertainties well below \(0.5^\circ\), high-energy tracks face challenges in energy reconstruction~\cite{IceCube:2013dkx}. Unless a muon track is both starting and stopping inside the detector, part of its energy loss occurs outside the instrumented volume and thus the muon energy at its production must be inferred. At high energies the process is nontrivial because the muon largely loses energy stochastically. Both the muon energy at production and the parent neutrino energy, which additionally depends on the inelasticity, are therefore dependent on the neutrino spectrum. The 2014 through-going track, shown in \cref{fig:evt1}, had a deposited energy inside IceCube of \SI{2.6(0.3)}{\peta \eV}, corresponding to a reconstructed muon energy of \SI{4.5(1.3:1.2)}{\peta \eV} at detector entry. Adopting an \(E^{-2.13}\) spectrum, the most probable neutrino energy was inferred to be \(8.7\)~PeV.

Tracks with an interaction vertex inside the instrumented volume, although missing a fraction of their energy once the track exits the detector, benefit from the fact that the first interaction products from DIS---including those from hadronization---are visible. Consequently, the inferred neutrino energy is less dependent on the assumed astrophysical flux prior and carries smaller uncertainties. The event IC-20190331A, show in \cref{fig:evt3}, was detected as a real-time alert~\cite{IceCube2019_GCN24028} and was also selected in both the HESE~\cite{IceCube:2023sov} and MESE~\cite{Basu:2025vbm} samples. Sophisticated reconstruction techniques were applied, matching the observed energy-loss profile to MC simulations. The average \(\dd E/\dd x\) over the last 400~m of the track was measured to be around \SI{1.125}{\tera \eV \per \m}, with a total visible energy of \SI{4.8}{\peta \eV}. Assuming the best-fit broken power-law spectrum from the MESE analysis~\cite{IceCube:2025tgp}, the inferred neutrino energy corresponds to \SI{11.4(2.46:2.53)}{\peta \eV} (c.f.~\cref{fig:evt3_energy}).

Shower events achieve an energy resolution at the $\sim 10\%$ level, since most of the secondary particle energy is deposited within the detector, allowing for a calorimetric measurement. Although the Cherenkov light from individual charged particles in the cascade is emitted along the Cherenkov angle of $\sim 41^\circ$, multiple scattering of photons in the ice smears out this initial anisotropy. At typical DOM distances of $\mathcal{O}(30$–$150)\,\mathrm{m}$ from the shower vertex, the photon traverses several scattering lengths (\SIrange{20}{50}{\m}). Even when a shower is only partially contained, the detected photons can be used to reconstruct the direction and the total deposited energy of the shower. In IceCube, maximum-likelihood reconstruction techniques are used to fit the event parameters $(x, y, z, t, \theta, \phi)$ by comparing the observed photon arrival times against expectations from approximate photon-yield models using splines~\cite{IceCube:2013dkx,IceCube:2024csv} or neural networks~\cite{Abbasi:2021ryj}, or fully resimulated photon propagation methods~\cite{Chirkin:2013avz}.

The partially contained shower shown in \cref{fig:evt2} is IceCube’s first Glashow resonance candidate. Detected on 2016-12-08 01{:}47{:}59~UTC, the event has a reconstructed visible energy of \SI{6.05(0.72)}{\peta \eV}~\cite{IceCube:2021rpz}. This is consistent with the resonant interaction
\begin{equation}
\bar{\nu}_e + e^- \rightarrow W^-
\label{eq:inverse_beta}
\end{equation}
at $E_{\bar{\nu}_e} \simeq \SI{6.3}{\peta \eV}$, followed by hadronic decay of the \(W^-\) boson. For such hadronic decays, simulations at \SI{6.3}{\peta \eV} predict that roughly \(5\%\) of the total energy is carried away by neutral particles or other components below the Cherenkov detection threshold. Applying this expected correction to the visible energy yields a total neutrino energy consistent with the Glashow resonance value of $\sim \SI{6.3}{\peta \eV}$.

In hadronic \(W^-\) decays, pions, kaons, and other mesons are produced, many of which subsequently decay into secondary muons. The main particle shower develops over \(\mathcal{O}(10\text{–}20)\,\mathrm{m}\), while the resulting Cherenkov photons propagate at $c/n < c$, where $n$ is the group index of refraction in the medium. Additionally, in-ice scattering can broaden and delay the photon arrival-time distribution. In contrast, the secondary muons are typically minimum-ionizing and travels, at $\sim c$, tens to hundreds of meters in ice, emitting Cherenkov light along their tracks. When such a muon passes closer to an optical module than the cascade vertex, its promptly emitted, weakly scattered light can produce ``early'' hits that precede the bulk of the scattered cascade light arriving at that sensor.

The full-waveform digitization of IceCube’s readout and \(\sim 2\text{--}3\) ns timing resolution per DOM allowed for distinctive early pulses to be observed on sensors closest to the reconstructed vertex of the Glashow resonance candidate, as shown in \cref{fig:waveform}. These early hits are naturally explained by secondary muons, with a most-probable leading-muon energy of \(\SI{26.4(28.6:12.4)}{\giga \eV}\), produced in the hadronic \(W^-\) decay and outrunning the Cherenkov wavefront of the primary hadronic shower, as illustrated in \cref{fig:rect1}. Further, a hybrid reconstruction that combines the standard cascade vertex fit with a muon-like subtrack improves the directional localization by about a factor of five (in area) compared to the cascade-only reconstruction, and is also less sensitive to systematic uncertainties in ice modeling.

All three neutrino events selected in the EHE analysis are more consistent with an astrophysical spectrum than with a cosmogenic origin~\cite{IceCube:2025ezc}. Consequently, an upper limit on the cosmogenic neutrino flux was derived, shown as the blue dashed line in Fig.~\ref{fig:overview_diffuse}. The mild excess around \(10~\mathrm{PeV}\) in the limit curve arises because three measured events fall in this energy range, making the observed limit weaker than the sensitivity. Future iterations of the IceCube EHE analysis~\cite{Nakos:2025yqb} will incorporate graph neural networks and transformer architectures to improve cosmic-ray background rejection. In particular, these methods will exploit information on muon bundles: the lateral separation of muons within a bundle produces characteristic hit patterns in the detector that differ from those of a single, isolated muon track. A global fit is also planned to combine constraints from atmospheric, astrophysical, and cosmogenic neutrino measurements, with a consistent treatment of detector and atmospheric systematics.

\subsection{Results of EHE neutrino observations from KM3NeT}

Data from 287.4~days of livetime, collected between 23~September~2022 and 11~September~2023 with the 21-line detector configuration, were analysed after removing periods used for detector commissioning and calibration~\cite{KM3NeT:2025npi}. During this time, about \num{1.1e8} events were triggered, corresponding to an average trigger rate of $\sim \SI{4.4}{\Hz}$. The brightest event in this dataset was KM3-230213A, which triggered 3,672~PMTs---about $35\%$ of the active PMTs in the 21-line array.

\begin{figure}[htbp]
  \centering
  \includegraphics[width=0.40\textwidth]{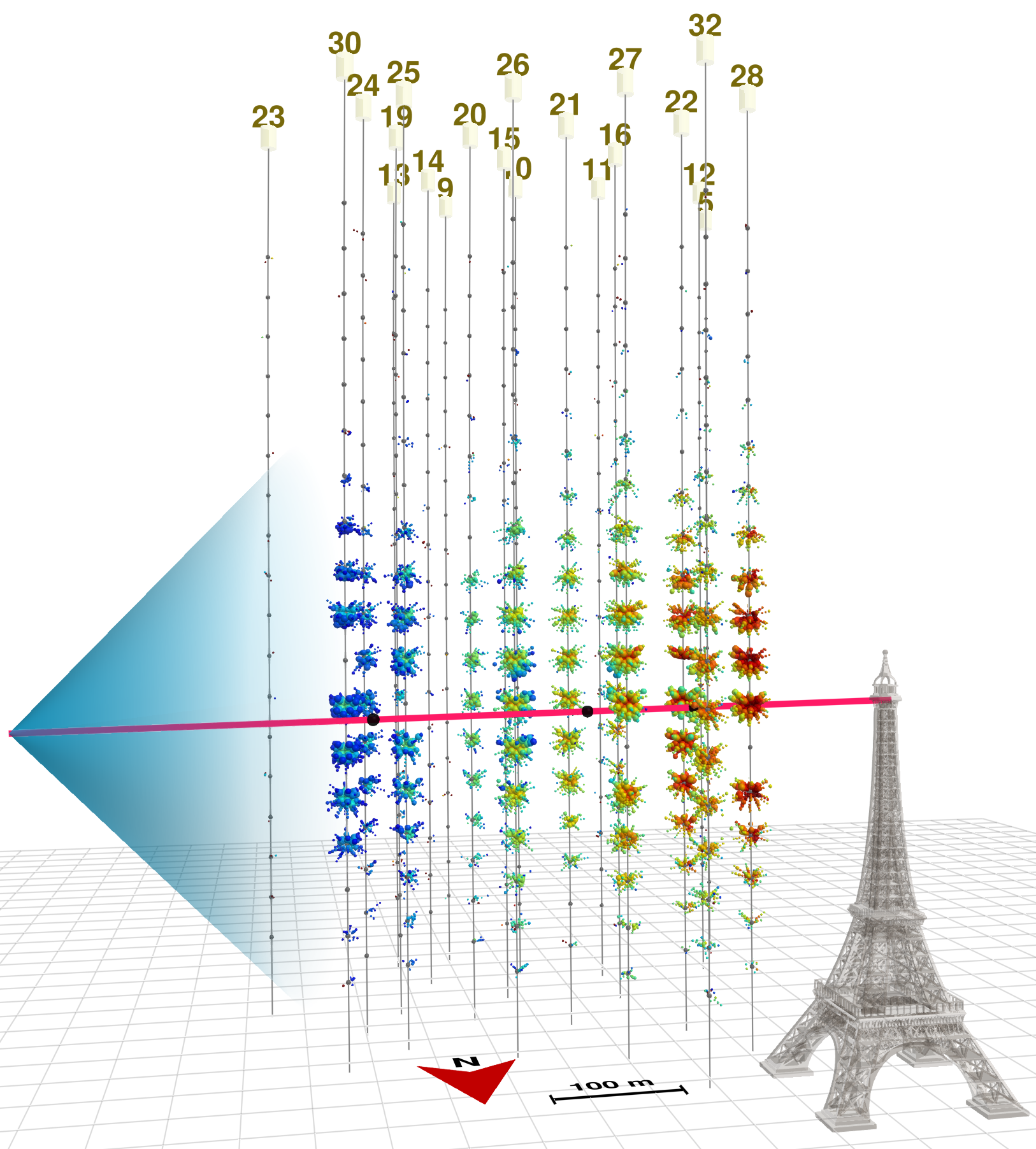}
  \hfill
  \includegraphics[width=0.55\textwidth]{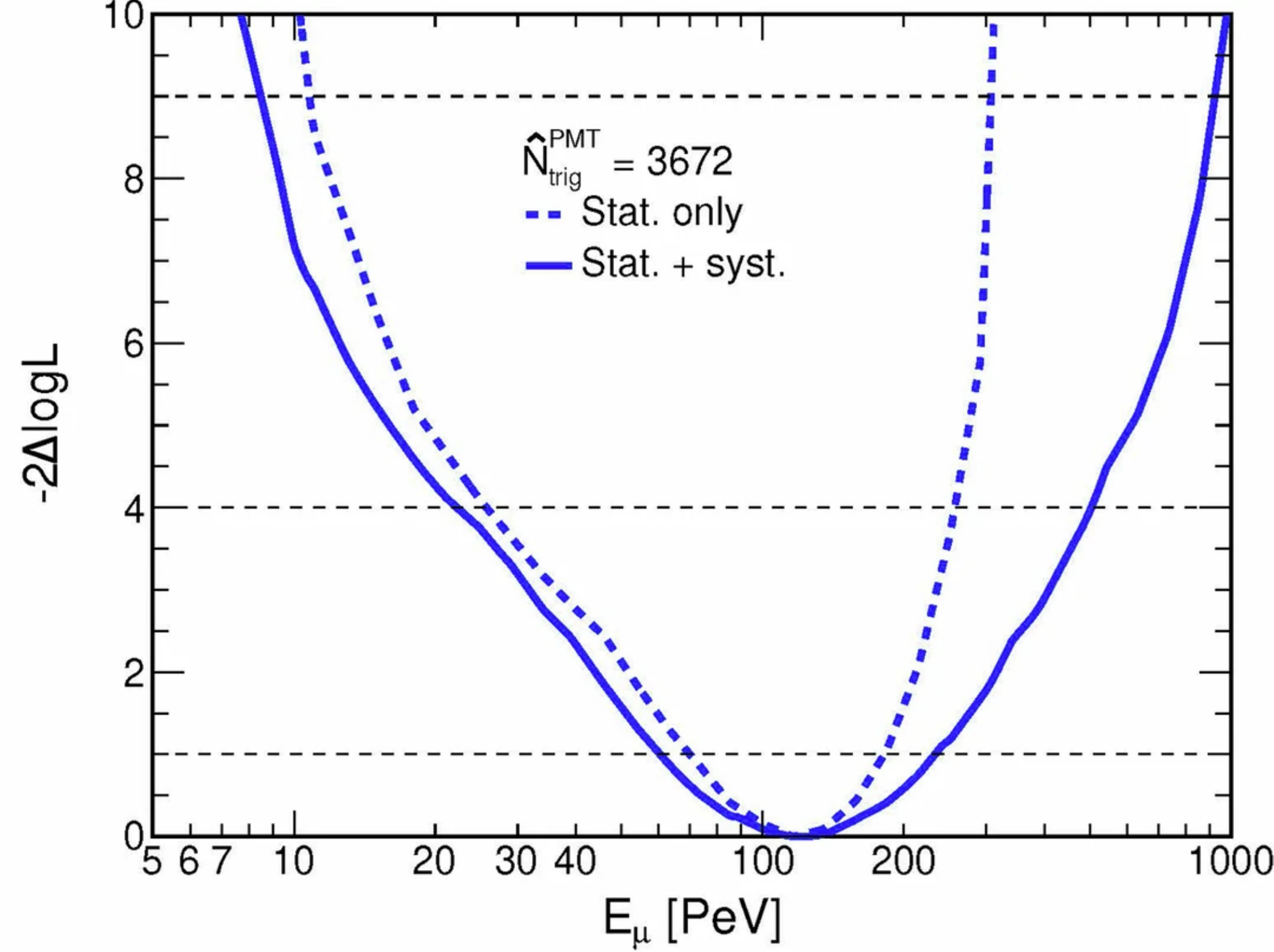}
  \caption{
  The KM3-230213A event. 
  Left: Event view of KM3-230213A, produced using the RainbowAlga event display software released with~\cite{KM3NeT:2025npi}. 
  The marker orientation indicates the sensitive direction of each PMT within a DOM, and the colors represent photon arrival times. 
  Right: Reconstructed muon energy likelihood based on the number of PMTs registering hits~\cite{KM3NeT:2025npi}. 
  Systematic uncertainties account for a 10\% variation in the assumed photon absorption properties of seawater. 
  No prior is applied on the astrophysical neutrino flux, and all energies are assumed to be equally probable for an astrophysical origin.
  }
  \label{fig:km3_evt_likelihood}
\end{figure}

The number of triggered PMTs, $\npmt$, was used as a proxy for reconstructing the muon energy \(E_{\mu}\).
MC simulations were generated for muons traversing the detector along the reconstructed direction of the event, at \emph{discrete} true muon energies \(E_{\mu}^{\rm true} \in \{1, 2,\,\ldots,\,1000\}\ \mathrm{PeV}\), and for multiple systematic configurations, including \(\pm 10\%\) variations of optical absorption and detection efficiency.  
For each grid point, a probability distribution \(P(\npmtnohat \mid E_{\mu}, \vec{\xi})\) was obtained, where $\npmtnohat$ is the number of triggered PMTs in the simulation and \(\vec{\xi}\) denotes the nuisance parameters. The data-constrained likelihood function
\begin{equation}
\mathcal{L}(E_{\mu},\vec{\xi}) =
P\big(\npmt \,\big|\, E_{\mu}, \vec{\xi}\big) \times
\prod_{i} \exp\!\left[
    -\frac{(\xi_i - \xi_{i,0})^2}{2\sigma_{\xi_i}^2}
\right],
\label{eq:likelihood}
\end{equation}
where $\npmtnohat$ is the number of triggered PMTs, \(\xi_{i,0}\) the nominal value of each nuisance parameter, and \(\sigma_{\xi_i} = 0.1\,\xi_{i,0}\) encodes the \(\pm 10\%\) Gaussian prior. The most-likely muon energy estimate corresponds to the \(E_{\mu}\) that maximizes \(\mathcal{L}\). The profile likelihood as a function of the muon energy, \(E_{\mu}\), is shown in the right panel of \cref{fig:km3_evt_likelihood}, yielding a best-fit value of \(E_{\mu} = \SI{110}{PeV}\). The \(\SI{1}{\sigma}\) confidence interval, \(E_{\mu} = 110^{+60}_{-30}~\mathrm{PeV}\), is derived according to Wilks' theorem with \(-2\,\Delta\ln\mathcal{L} = 1\) for one degree of freedom.

\begin{figure}[htbp]
  \centering
  \includegraphics[width=0.8\textwidth]{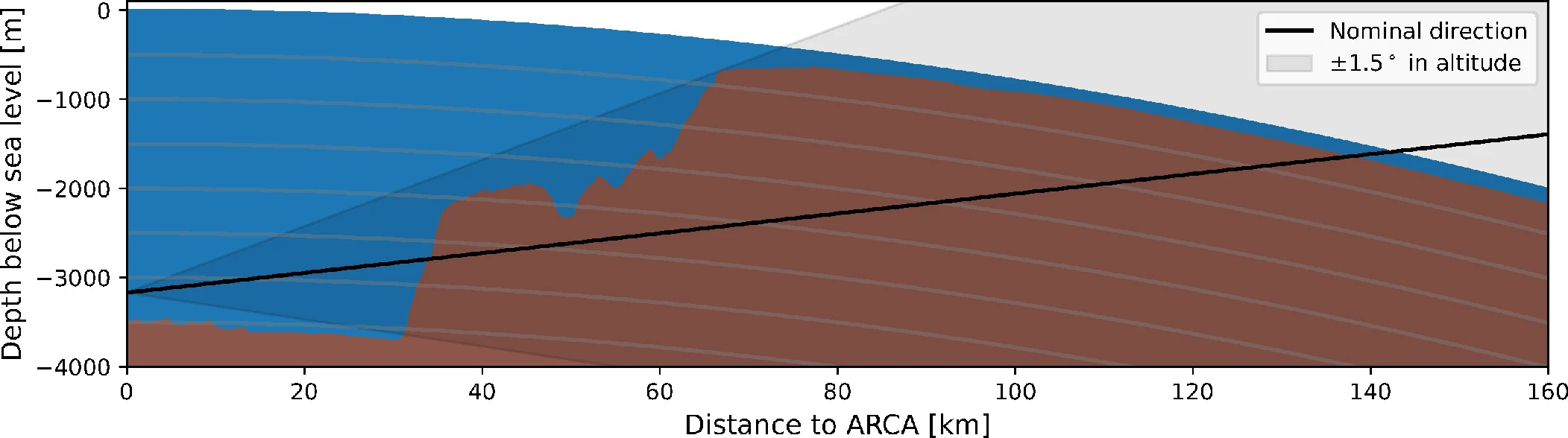}
    \caption{
    Rock and water overburden above the KM3NeT detector for the KM3-230213A event~\cite{KM3NeT:2025npi}. 
    The solid line shows the best-fit reconstructed direction, and the shaded area indicates the 1$\sigma$ angular uncertainty of 1.5$^\circ$. 
    Blue shading corresponds to seawater, while brown represents the seafloor and underlying rock.
    }

  \label{fig:km3_overburden}
\end{figure}
The reconstruction assumes the event is a single muon, not a muon bundle from a cosmic-ray air shower. This is because the inclined zenith angle of \(89.4^\circ\) corresponds to a slant depth of \(\sim 300~\mathrm{km}\) water equivalent along the reconstructed trajectory, and passes through the Malta escarpment as shown in \cref{fig:km3_overburden}. As a reference typically \(\sim \SI{100}{\exa \eV}\) muons have average range of \(\sim \SI{60}{\kilo \m}\) water equivalent. Dedicated simulations of atmospheric muons, incorporating a zenith-angle uncertainty of up to \(2^\circ\), yield an upper limit on the background rate of \(\sim 10^{-10}~\mathrm{yr}^{-1}\) at \(E_\mu \simeq 100~\mathrm{PeV}\), increasing to \(\sim 10^{-9}~\mathrm{yr}^{-1}\) for \(E_\mu \simeq 10~\mathrm{PeV}\)~\cite{KM3NeT:2025npi}. In an extremely unlikely scenario where the reconstructed zenith angle were biased by \(\SI{5}{\sigma}\) (\ang{5.6} above the horizon), the slant depth would be reduced to \(\sim \SI{28}{\kilo m}\) water equivalent, corresponding to an atmospheric muon background expectation of order \(10^{-3}\)~events per year.

Although the statistical uncertainty in the reconstructed direction of this muon is \(0.12^\circ\), the current systematic uncertainty on the absolute orientation of the detector on the Earth results in an estimated angular uncertainty of \(\sim \ang{1.5}\) at the \(68\%\) confidence level, increasing to \(\sim \ang{3}\) at the \(99\%\) level. A recalibration campaign~\cite{Dornic2025_KM3NeTStatus} has been proposed to improve the absolute positioning of the DUs on the seafloor, thereby reducing these systematic uncertainties.

\begin{figure}[htbp]
  \centering
  \includegraphics[width=0.85\textwidth]{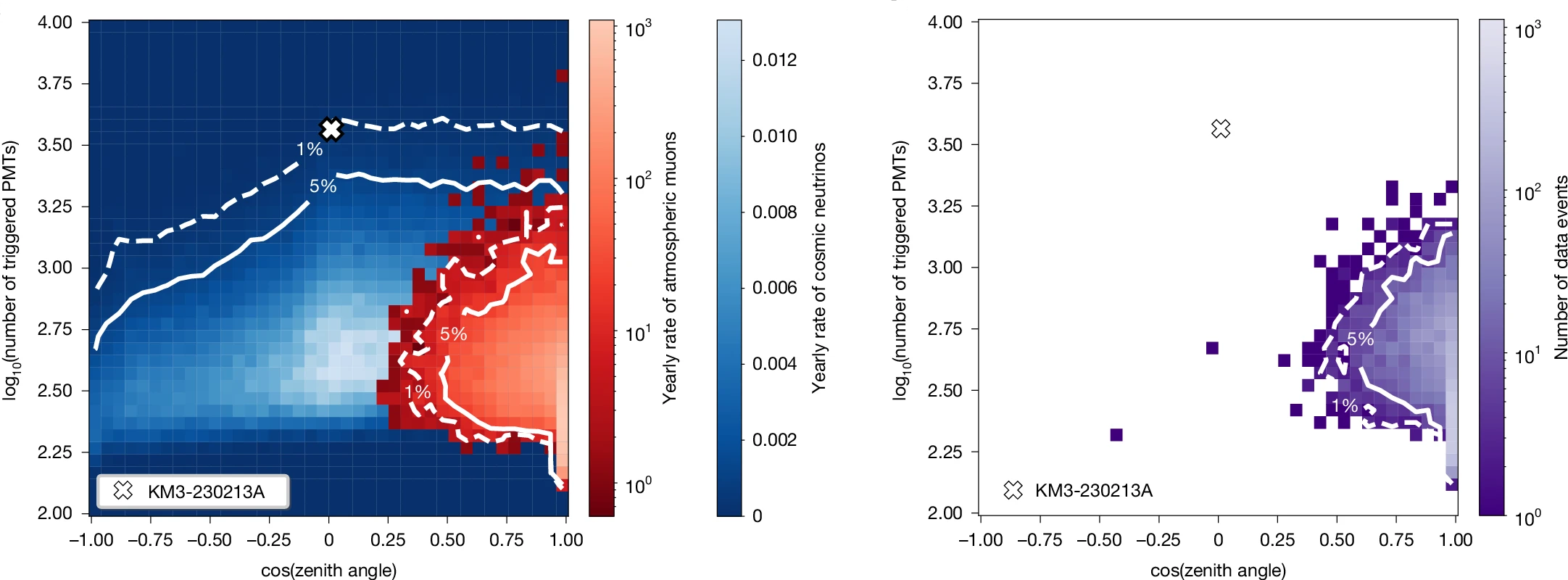}
    \caption{
    Number of triggered PMTs versus $\cos(\theta_{\mathrm{zenith}})$ from simulation (left) and data (right) for the high-quality track selection~\cite{KM3NeT:2025npi}. In both panels, the cross marks KM3-230213A.
    Left: Red hues indicate atmospheric muon background, concentrated in the down-going direction, while blue hues show astrophysical neutrinos simulated with a single power-law spectrum of index 2.37~\cite{Abbasi:2021qfz}. 
    Right: Purple hues represent the data distribution with the same selection. The two other outlier events are up-going with lower energies, and could originate from either astrophysical or atmospheric neutrinos.
    }

  \label{fig:km3_bkg}
\end{figure}

A comparison of data and MC distributions of the number of triggered PMTs versus \(\cos\theta\) is shown in \cref{fig:km3_bkg}, where \(\theta\) is the reconstructed zenith angle. The cosmic-ray background is concentrated in the down-going region; notably, the most horizontal bin where there is background corresponds to \(\cos\theta \approx 0.25\), or a zenith angle of \(\theta \approx \ang{75.5}\). Although high-$\npmtnohat$ background events are present in the MC, they are predominantly down-going. The simulated cosmic-neutrino sample is weighted according to a SPL astrophysical flux with spectral index \(\gamma = 2.37\), and KM3-230213A is marked by a cross. The dashed line indicates the region outside of which fewer than \(1\%\) of events from either background or the simulated astrophysical flux are expected, highlighting that KM3-230213A is in tension with the SPL flux measured by IceCube. \Cref{fig:km3_bkg} also shows the data distribution in the right panel. More detailed discussions in terms of consistencies with IceCube will be presented in the next section.

We note that the reconstruction of this event assumes the observed track was produced by a muon rather than a tau lepton. In general, taus lose energy more slowly than muons, with the average energy loss per unit column depth expressed as
\begin{equation}
-\frac{\dd E}{\dd X} = \alpha + \beta E,
\label{eq:energy_loss}
\end{equation}
where \(\alpha\) represents the ionization term and \(\beta\) accounts for the sum of radiative losses (bremsstrahlung, pair production, photonuclear interactions). Simulations with PROPOSAL~\cite{Dunsch:2018nsc} show that at PeV energies in water or ice, \(\beta_{\tau}\) is significantly smaller than \(\beta_{\mu}\), due to the tau’s larger mass and suppressed radiative cross sections. This implies that, for the same track length and observed light yield, a tau lepton would have had a substantially higher initial energy than a muon. A notable example is the IceCube event IC-20140611, which has been hypothesised to originate from a \(\nu_\tau\) with energy well above \(10~\mathrm{PeV}\), potentially approaching \(\mathcal{O}(100~\mathrm{PeV})\)~\cite{Kistler:2017xx}. In that scenario, the tau either traversed the detector directly or produced a muon via decay outside the instrumented volume, leading to a much larger parent neutrino energy than inferred under the muon assumption.

A summary of neutrino candidates above \SI{5}{\peta \eV} is provided in \cref{tab:uhe_events}. We note, however, that there exists an additional IceCube event above this threshold which is not included in this review. This event was originally reported in an analysis employing IceTop as a veto in combination with a data-driven background estimation~\cite{Lyu:2024jsm}. The inferred muon energy was \SI{8.1}{\peta \eV} with \(\cos\theta_{\mathrm{zen}} = 0.74\), and the reported signalness was \(1.0 \pm 0.47\). Due to the large uncertainty in the signalness, and the fact that a more sophisticated energy reconstruction would be required to convert the truncated-energy proxy to a neutrino energy, this event is not listed in \cref{tab:uhe_events} as part of this review.

The current IceCube EHE upper limit~\cite{IceCube:2025ezc} is shown in \cref{fig:overview_diffuse} as the blue, dashed line, which can be compared against the KM3NeT EHE flux shown as the purple error bar~\cite{KM3NeT:2025npi} in the same figure. Assuming an $E^{-2}$ shape, the best-fit flux normalization from KM3NeT is given as,
\begin{equation}
E^{2} \Phi = \SI{5.8e-8}{\giga \eV \per \centi \m \squared \per \s \per \steradian},
\label{eq:flux}
\end{equation}
in tension with current IceCube \SI{90}{\%} upper limits. However, this is a standalone interpretation of KM3-230213A using only KM3NeT/ARCA21. As will be discussed, a combined global analysis incorporating all experiments substantially reduces this apparent tension.  

\begin{table}[htbp]
\centering
\caption{Summary of the highest-energy neutrino candidate events detected by IceCube and KM3NeT. Energies are reconstructed or inferred assuming the quoted spectral priors.}
\label{tab:uhe_events}
\resizebox{\textwidth}{!}{%
\begin{tabular}{lccccl}
\hline
Event & Date (UTC) & Detector & Visible Energy [PeV] & Inferred $E_\nu$ [PeV] & Notes \\
\hline
IC-20140611 & 2014-06-11 04:54:24 & IceCube IC86 & $2.6 \pm 0.3$ & $8.7$ (E$^{-2.13}$ prior) & Through-going track \\
IC-20161208 & 2016-12-08 01:47:59 & IceCube IC86 & $6.05 \pm 0.72$ & $\sim 6.3$ & Partially contained shower \\
IC-20190331A & 2019-03-31 06:55:43 & IceCube IC86 & $4.8$ & $11.4^{+2.46}_{-2.53}$ & Starting track \\
KM3-230213A & 2023-02-13 01:16:47 & KM3NeT ARCA21 & --- & $220^{+570}_{-110}$ (E$^{-2}$ prior) & Through-going track \\
\hline
\end{tabular}%
}
\end{table}

\section{Interpretation and discussion}
\label{sec:disc}

In light of the recent KM3NeT/ARCA detection of what is likely the highest-energy neutrino candidate, it is important to assess the result in context within the global neutrino landscape. With only a handful of events above \SI{5}{\peta \eV} detected by IceCube, and none other above roughly \SI{20}{\peta \eV}, the natural inclination is to attempt to reconcile prior experimental results with the $\mathcal{O}(\SI{100}{\peta \eV})$ neutrino candidate KM3-230213A. In~\cite{KM3NeT:2025ccp}, the KM3NeT collaboration provides an in-depth study of different scenarios with different sample combinations from IceCube, Auger and KM3NeT. They find a tension in the range of \SIrange{2.5}{3}{$\sigma$}, with a few scenarios where high-energy IceCube measurements are included lowering the tension to about \SI{1.6}{$\sigma$}. Separately, an independent team finds a \SI{3.5}{$\sigma$} tension between the IceCube and KM3NeT data under a diffuse, isotropic flux assumption~\cite{Li:2025tqf}. The least disfavored scenario is that of a transient source, which is suggested to lower the tension to \SI{2}{$\sigma$}, although it is worth noting that this requires specific assumptions on the transient model. Along the same vein, Ref.~\cite{Das:2025vqd} suggests a new transient population that is energetic and $\gamma$-ray dark could explain the origins of KM3-230213A, although it is worth noting that such a population of sources likely should have also been detected by IceCube over its longer livetime. Additionally, Ref.~\cite{Muzio:2025gbr} performed a diffuse flux measurement based on all available neutrino data above \SI{5}{\peta \eV}, combined with Auger CR spectrum and composition data, for distinct energy scenarios of KM3-230213A. In the \SI{100}{\peta \eV} scenario the best-fit flux lies below the latest IceCube EHE limits~\cite{IceCube:2025ezc}, while in the \SI{1}{\exa \eV} scenario the measured flux is consistent with the EHE limit at the \SI{1}{$\sigma$} level. Indeed, the cosmogenic origin interpretation has been discussed by numerous authors~\cite{Muzio:2025gbr,Das:2020nvx,Kuznetsov:2025ehl,Zhang:2025abk}, as well as the KM3NeT collaboration itself~\cite{KM3NeT:2025vut}. A crucial component to any interpretation is the energy estimator. In this \namecref{sec:disc}, we first provide an in-depth discussion of the inferred neutrino energy of KM3-230213A, followed by a review of the various consistency checks that have been performed with the global neutrino data.

\subsection{Neutrino energy estimation of KM3-230213A}
\label{sec:enu}

Under the assumptions that KM3-230213A was due to a CC DIS muon neutrino interaction on an $E_\nu ^{-2}$ flux, the KM3NeT collaboration quotes a median neutrino energy of \SI{220}{\peta \eV} with the 68\% interval spanning \SIrange{110}{790}{\peta \eV}~\cite{KM3NeT:2025npi}. While that reference does not provide energy estimates under alternative scenarios, it does provide the muon-energy likelihood, $\mathcal{L}(E_\mu | \npmt)$, given the number of triggered PMTs, $\npmt$ (c.f.~\cref{fig:km3_evt_likelihood}). The likelihood is constructed by simulating muons of various energies at the same position and direction as the event, and matching the number of triggered PMTs in simulation to data. A simple application of Bayes' theorem yields
\begin{equation} \label{eq:bayes}
    f(E_\nu|\npmt) = C \int \mathcal{L}(E_\mu| \npmt) f(E_\mu | E_\nu) f(E_\nu) \dd E_\mu,
\end{equation}
where $f(E_\nu | \npmt)$ is the posterior neutrino energy distribution, $C$ is a normalization constant, $f(E_\nu)$ the neutrino-energy prior, and $f(E_\mu | E_\nu)$ the conditional probability to observe a muon at the detector of energy $E_\mu$ given a muon neutrino of energy $E_\nu$ at Earth's surface. To evaluate $f(E_\mu | E_\nu)$, we used SIREN~\cite{Schneider:2024eej} to inject muon neutrinos along the best-fit zenith direction of \ang{89.4} towards the ARCA21 geometry. Secondary muons were then propagated with MMC~\cite{Chirkin:2004hz}, starting from the neutrino interaction vertex, to a plane placed at a depth of \SI{3.2}{\kilo \m} below sea level.

\begin{figure}[htbp]
  \centering
    \includegraphics[width=0.48\linewidth]{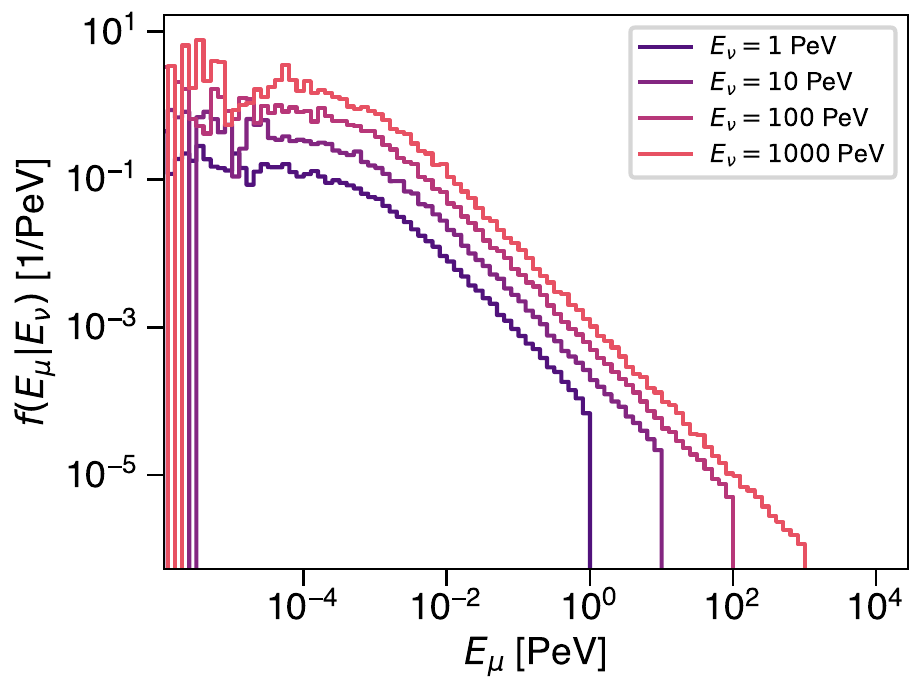}
    \includegraphics[width=0.48\linewidth]{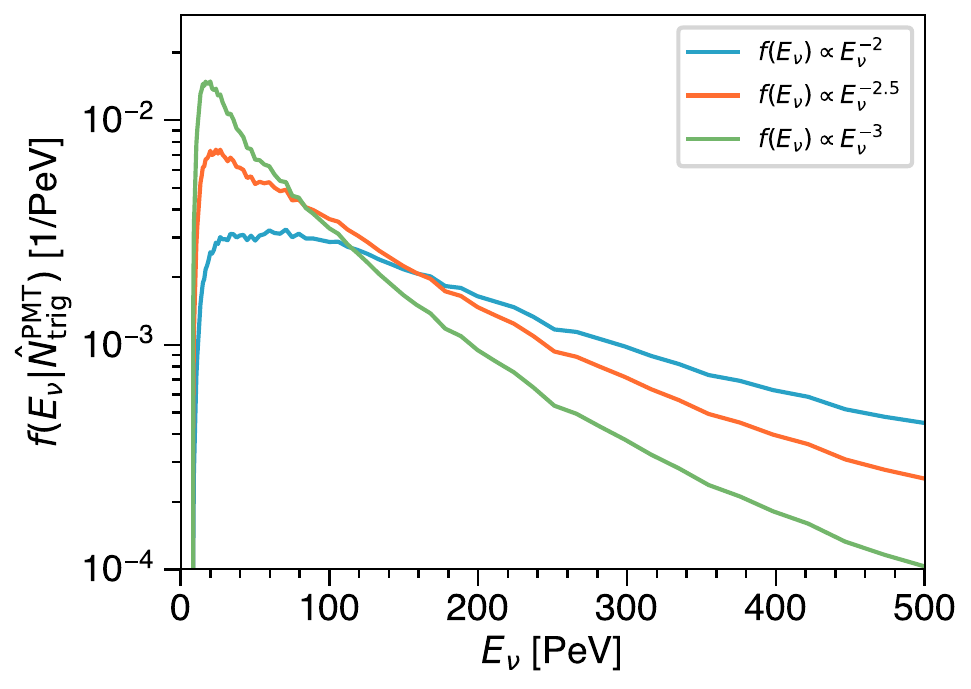}
  \caption{The left panel shows the probability density, $f(E_\mu | E_\nu)$, of a muon produced in CC DIS muon-neutrino scattering to arrive at the ARCA21 detector with energy $E_\mu$ for the neutrino energies indicated in the legend. The conditional probabilities are constructed using SIREN~\cite{Schneider:2024eej} as a neutrino generator and MMC~\cite{Chirkin:2004hz} as a muon propagator. For ease of visualization, only a subset of the simulated $E_\nu$ values are shown here. The right panel shows the posterior $f(E_\nu|\npmt)$ as evaluated based on \cref{eq:bayes} for the spectral priors shown in the legend.}
  \label{fig:fen}
\end{figure}
The left panel of \cref{fig:fen} shows the conditional probability, $f(E_\mu|E_\nu)$, for different values of $E_\nu$ each corresponding to a different colored line\footnote{In practice this is implemented as a transfer matrix.}. The injected neutrino energy corresponds to the upper bound of each distribution. The right panel of \cref{fig:fen} shows $f(E_\nu | \npmt)$ as evaluated by \cref{eq:bayes} for a few different spectral assumptions. The mode of each distribution is given in the legend. The long tail of the distributions implies that the median is pushed to higher energies, and we find that the 16--50--84 percent quantiles correspond to the values given in \cref{tab:percentiles}. For $\gamma=2$, the median $E_\nu$ agrees with the published value, though the 16\% and 84\% values differ somewhat with the 68\% intervals~\footnote{The exact quantiles are not specified in Ref.~\cite{KM3NeT:2025npi}.} in~\cite{KM3NeT:2025npi}. One difference is that in Ref.~\cite{KM3NeT:2025npi}, even though the muon energy estimate is conditioned on the event position and direction, the neutrino energy estimate is based on neutrinos simulated from the entire sky.  Here we have constrained the neutrino direction along the reported best-fit zenith and azimuth. Along that direction, the Malta escarpment protrudes above the seafloor, as shown in Extended Data Fig.~4 in \cite{KM3NeT:2025npi}, at a distance of which we approximate to be about \SI{34}{\kilo \m} from the detector and account for in our MMC simulations. Another possible difference lies in the definition of where $E_\mu$ is defined. Here, it is taken to be at the surface of the smallest cylinder that surrounds the ARCA21 detection lines.
\begin{table}[ht]
\caption{The 16--50--84 percent quantiles of $f(E_\nu|\npmt)$ for three different power-law spectrum priors, $f(E_\nu)$, corresponding to those shown in the right panel of \cref{fig:fen}. For $\gamma=2$ the median $E_\nu$ agrees with the value quoted in~\cite{KM3NeT:2025npi}.}
\label{tab:percentiles}
\centering
\begin{tabular}{@{}lrrrr@{}}
\toprule
Spectral index ($\gamma$) & Mode [PeV] & 16\%  [PeV] & 50\% [PeV] & 84\% [PeV] \\
\midrule
2    & 70 & 70 & 225 & 890 \\
2.5  & 25 & 35 & 110 & 355 \\
3    & 20 & 20 & 60 & 180 \\
\bottomrule
\end{tabular}
\end{table}

An independent estimation of the neutrino energy was given in~\cite{Li:2025tqf}. Performed prior to the publication of~\cite{KM3NeT:2025npi}, the calculation used MadGraph to simulate CCDIS neutrino interaction rates and a simplified detector model to construct $L(E_\mu|\npmt)$ based on PROPOSAL simulations of the muon energy losses inside the detector. The posterior was then obtained with Bayes' theorem as in \cref{eq:bayes}. The 90\% credible intervals were found to be \SIrange{4}{760}{\peta \eV} (\SIrange{23}{2400}{\peta \eV}) for priors of $\gamma=2~(2.52)$, consistent with the results given here. A cosmogenic flux scenario from~\cite{Ahlers:2012rz} was also tested, resulting in 90\% intervals of \SIrange{93}{2400}{\peta \eV} for $E_\nu$. It follows that when inferring the neutrino energy of KM3-230213A the choice of prior should always be specified. 

\subsection{How consistent is KM3-230213A with IceCube results?}

Although the inferred neutrino energy of KM3-230213A comes with caveats of prior dependence and large uncertainties, odds are it is the highest-energy neutrino ever detected. Recent attempts have been made to interpret the data in both a standalone KM3NeT/ARCA context as well as a global dataset context including experimental observations and non-detections from IceCube and Auger. The diffuse flux reported in~\cite{KM3NeT:2025npi} was based on the standalone KM3NeT/ARCA exposure, and its best-fit (c.f.~\cref{fig:future_detectors}) lies an order of magnitude above the previous 90\% upper limits reported by IceCube~\cite{IceCube:2018fhm} and Auger~\cite{PierreAuger:2023pjg}, though with only one event the \SI{2}{$\sigma$} uncertainties extend below the IceCube limits. An independent analysis suggests that, assuming KM3-230213A arises from an $E^{-2.52}$ spectrum and the best-fit flux normalization from the KM3NeT/ARCA21 standalone result, IceCube should have seen 75 neutrinos in the \SIrange{72}{2600}{\peta \eV} energy range~\cite{Li:2025tqf}    . The discrepancy is driven by ARCA21's much smaller exposure—inversely proportional to the flux—compared to IceCube. While such KM3NeT/ARCA standalone analyses lead to strong tension with existing IceCube data, combining the ARCA21 result with IceCube and Auger non-observation decreases the best-fit flux substantially, with 0.59, 0.40, and 0.013 events expected in IceCube, Auger, and ARCA21, corresponding to a data p-value of about 0.5\% (\SI{2.6}{$\sigma$})~\cite{KM3NeT:2025npi}. Since the diffuse flux is assumed to be isotropic in space and time, correctly accounting for the full exposure across all operational neutrino telescopes is necessary when assuming Poisson statistics.

\begin{figure}[htbp]
  \centering
  \includegraphics[width=0.9\textwidth]{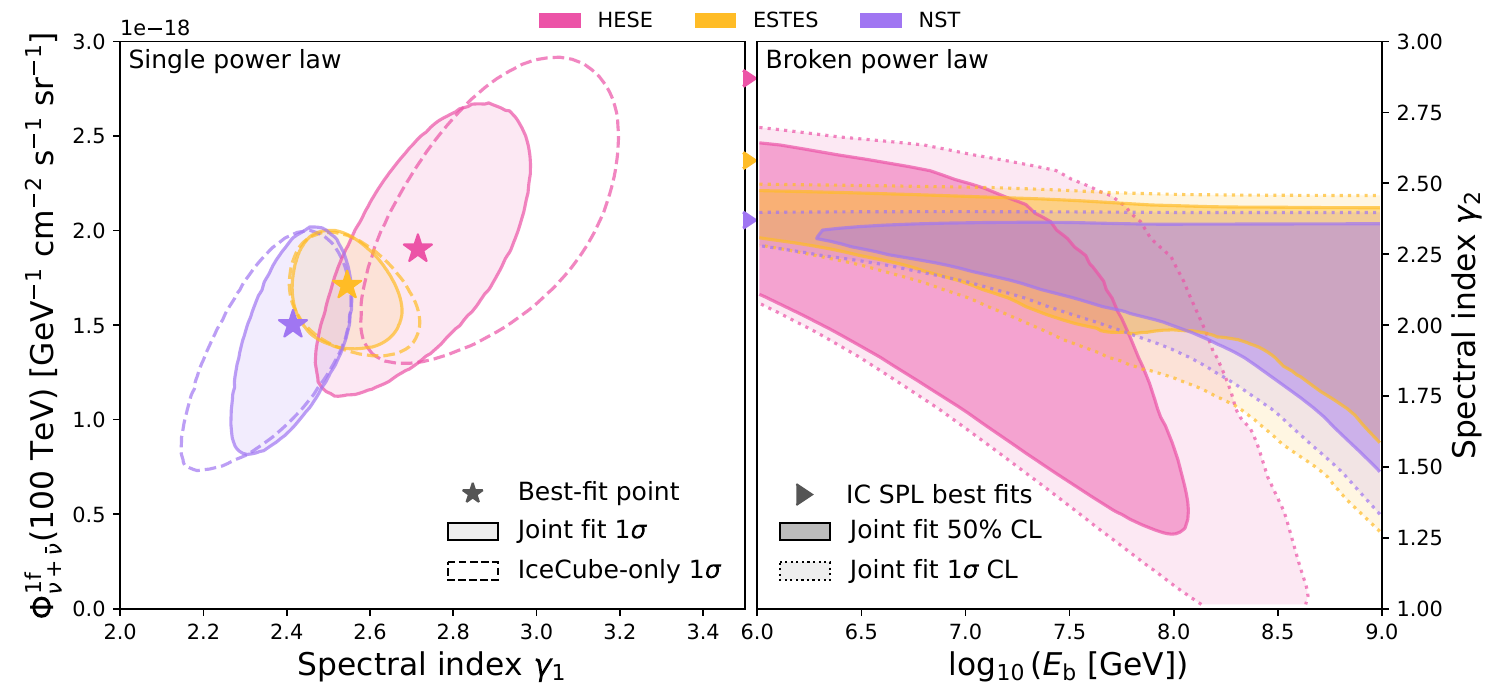}
  \caption{Figure from~\cite{KM3NeT:2025ccp}. The shaded regions in the right panel shows \SI{1}{$\sigma$} SPL contours from a joint analysis of KM3NeT, IceCube and Auger data at energies above tens of PeV in combination with one of the individual IceCube HESE (magenta), ESTES (yellow) or NT (purple) dataset. Dashed lines show the IceCube-only contours. The left panel shows results from a BPL fit of the break energy, $E_b$ and $\gamma_2$. The triangles indicate the respective IceCube SPL spectral index best-fits.}
  \label{fig:km3_jointfit}
\end{figure}
In a follow up publication, the KM3NeT collaboration reported on a combined IceCube, Auger and KM3NeT analysis~\cite{KM3NeT:2025ccp}. Motivated by their median $E_\nu$ of \SI{220}{\peta \eV} assuming an $E^{-2}$ spectrum, the analysis defines an EHE region above tens of \si{\peta \eV} where the only event detected is KM3-230213A. Non-observation of events by IceCube above and Auger are incorporated into a joint likelihood by taking the IceCube EHE exposure from~\cite{IceCube:2018fhm} and the Auger exposure from three different neutrino samples~\cite{PierreAuger:2019azx}. The IceCube EHE energy range is assumed to span \SI{20}{\peta \eV} to \SI{50}{\exa \eV}, while Auger's spans \SI{10}{\peta \eV} to \SI{1}{\zetta \eV}. Taking the one event detected across all three experiments, and assuming an $E^{-2}$ spectrum, the best-fit flux normalization is found to be compatible with existing limits from both IceCube and Auger. To quantify the tension betwen the datasets, a goodness-of-fit metric was constructed under a SPL hypothesis to be the ratio of best-fit likelihood under a joint-experiment fit to the best-fit likelihood when individual samples are independently fitted. A \SI{2.5}{$\sigma$} tension is reported~\cite{KM3NeT:2025ccp}.

In addition to the analysis involving only EHE samples across the three observatories, a combination of those with an additional IceCube high-energy (HE) sample chosen from the HESE~\cite{IceCube:2020wum}, ESTES~\cite{IceCube:2024fxo} and NT~\cite{Abbasi:2021qfz} analyses were performed in~\cite{KM3NeT:2025ccp}. To avoid double counting due to events shared across samples, the HE samples were tested one at a time. For simplicity, the SPL likelihood space from IceCube publications was directly added to form a joint EHE+HE likelihood. The impact on the SPL contours are shown in the left panel of \cref{fig:km3_jointfit}, where the dashed lines show the IceCube results while shaded regions of the same color highlight the joint EHE+HE result including KM3NeT, IceCube and Auger EHE datasets. A BPL hypothesis, with $E_b$ at energies above \SI{1}{\peta \eV} was also tested, marginalizing over $\Phi$ and  $\gamma_1$. The result is shown in the right panel of \cref{fig:km3_jointfit}.

\begin{figure}[htbp]
    \centering
    \includegraphics[width=0.48\linewidth] {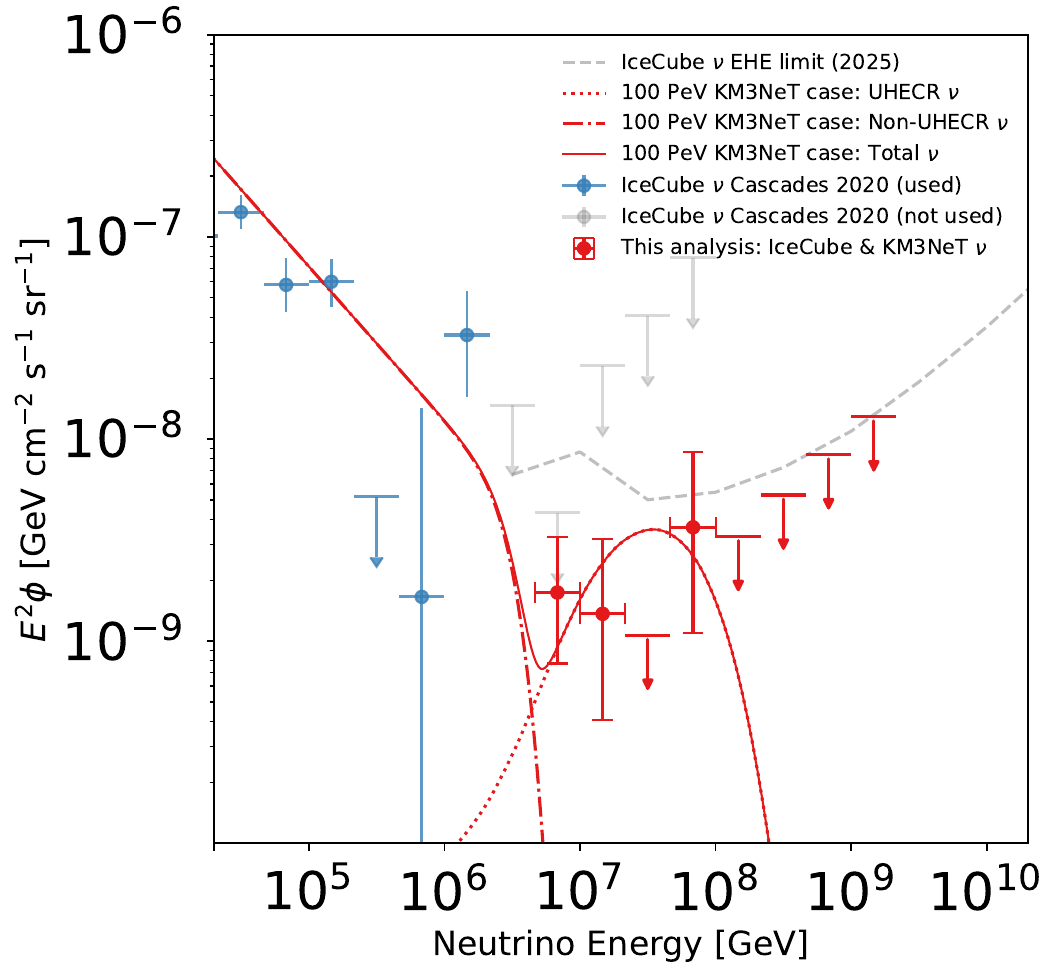} 
    \includegraphics[width=0.48\linewidth]{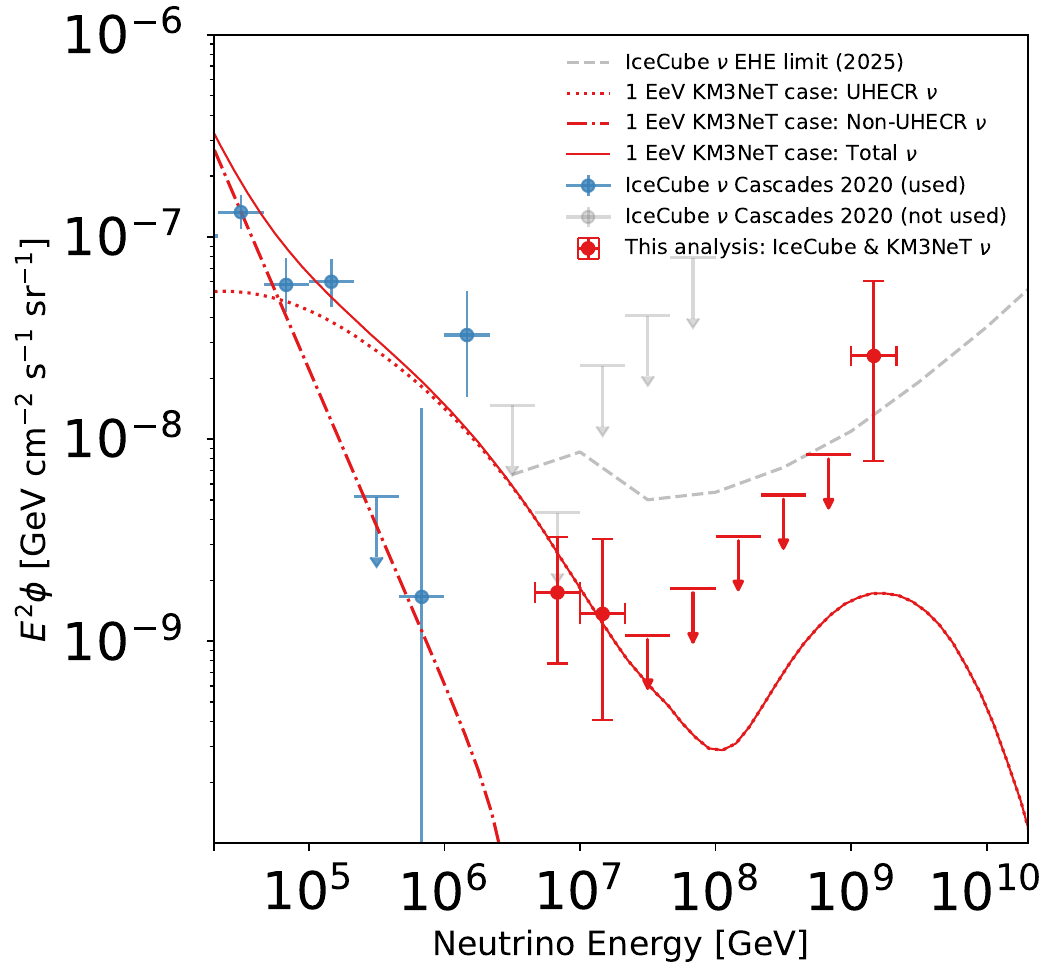} 
    \caption{The left (right) panel shows results obtained for the \SI{100}{\peta \eV} (\SI{1}{\exa \eV}) energy scenario of the KM3NeT/ARCA event. The results are shown in red, with error bars from a piecewise-segmented unfolding routine with each segment assumed to follow an $E^{-2}$ shape, and lines corresponding to predictions from the UFAM model that best fits the neutrino and CR data. The UHECR model in the right panel includes an additional, pure-proton population at the highest energies. Data points from Ref.~\cite{IceCube:2020acn} are shown in blue and grey, with blue points being included in the fit to constrain the lower energy regime. The dashed line corresponds to the 90\% upper limit from Ref.~\cite{IceCube:2025ezc}.}
    \label{fig:uhecrnu}
\end{figure}
Taking the global dataset one step further, Ref.~\cite{Muzio:2025gbr} performs a combined fit of neutrino data from KM3NeT and IceCube above \SI{5}{\peta \eV} with CR spectrum and composition data from Auger. The large $E_\nu$ uncertainty of KM3-230213A, in contrast to the relatively well constrained energies of IceCube neutrinos given in \cref{tab:uhe_events}, allowed for some flexibility in the event's interpretation with two KM3-230213A energy scenarios tested: \SI{100}{\peta \eV} and \SI{1}{\exa \eV}. The IceCube events above \SI{5}{\peta \eV} were reported in three independent and largely non-overlapping samples, HESE, NT and PEPE~\cite{Lu:2017nti}. This meant a joint likelihood could be constructed by using their respective effective areas and exposures and summing over the log-likelihoods of each sample. Since the results were obtained prior to the publication of~\cite{KM3NeT:2025npi}, an approximation of the KM3NeT/ARCA effective area and exposure was assumed. The Auger CR spectrum data from~\cite{PierreAuger:2021hun} and composition data from~\cite{PierreAuger:2023kjt} were additionally included. In order to make the connection to UHECR, the Unger-Farrar-Anchordoqui (UFA) CR source model~\cite{Unger:2015laa} as extended by Muzio in~\cite{Muzio:2019leu,Muzio:2021zud} was used to derive predictions of both neutrino and CR expectations. Finally, in order to constrain the non-UHECR contribution to the neutrino flux, which rises at lower energies, the flux measurements in Ref.~\cite{IceCube:2020acn} below \SI{5}{\peta \eV} were directly included via gamma distributions into the joint fit.

Results for the UFAM model fits are shown as red lines in the left (right) panel of \cref{fig:uhecrnu} for the \SI{100}{\peta \eV} (\SI{1}{\exa \eV}) KM3-230213A energy scenario. In the \SI{1}{\exa \eV} case, an extra pure-proton component is introduced to better describe the highest-energy data point. The recovery of the neutrino flux in the right panel starting at around \SI{100}{\peta \eV} is due the interaction of these protons off the CMB, thus allowing for the interpretation of the KM3NeT event as the first GZK neutrino in this scenario. The blue (grey) error bars show flux measurements from~\cite{IceCube:2020acn} below (above) \SI{5}{\peta \eV}, with the blue being included in the UFAM model fit to constrain the lower-energy non-UHECR component. Additionally, a model-independent measurement of the neutrino flux above \SI{5}{\peta \eV} is shown as the red error bars, in which only the neutrino data is used to fit piecewise segments of the flux normalization assuming an $E^{-2}$ spectrum. We note that, even in the \SI{1}{\exa \eV} scenario the measured flux is consistent with the 90\% IceCube upper limits~\cite{IceCube:2025ezc}, though not necessarily with the IceCube SPL best-fits.

\begin{figure}[htbp]
    \centering
    \includegraphics[width=0.48\linewidth] {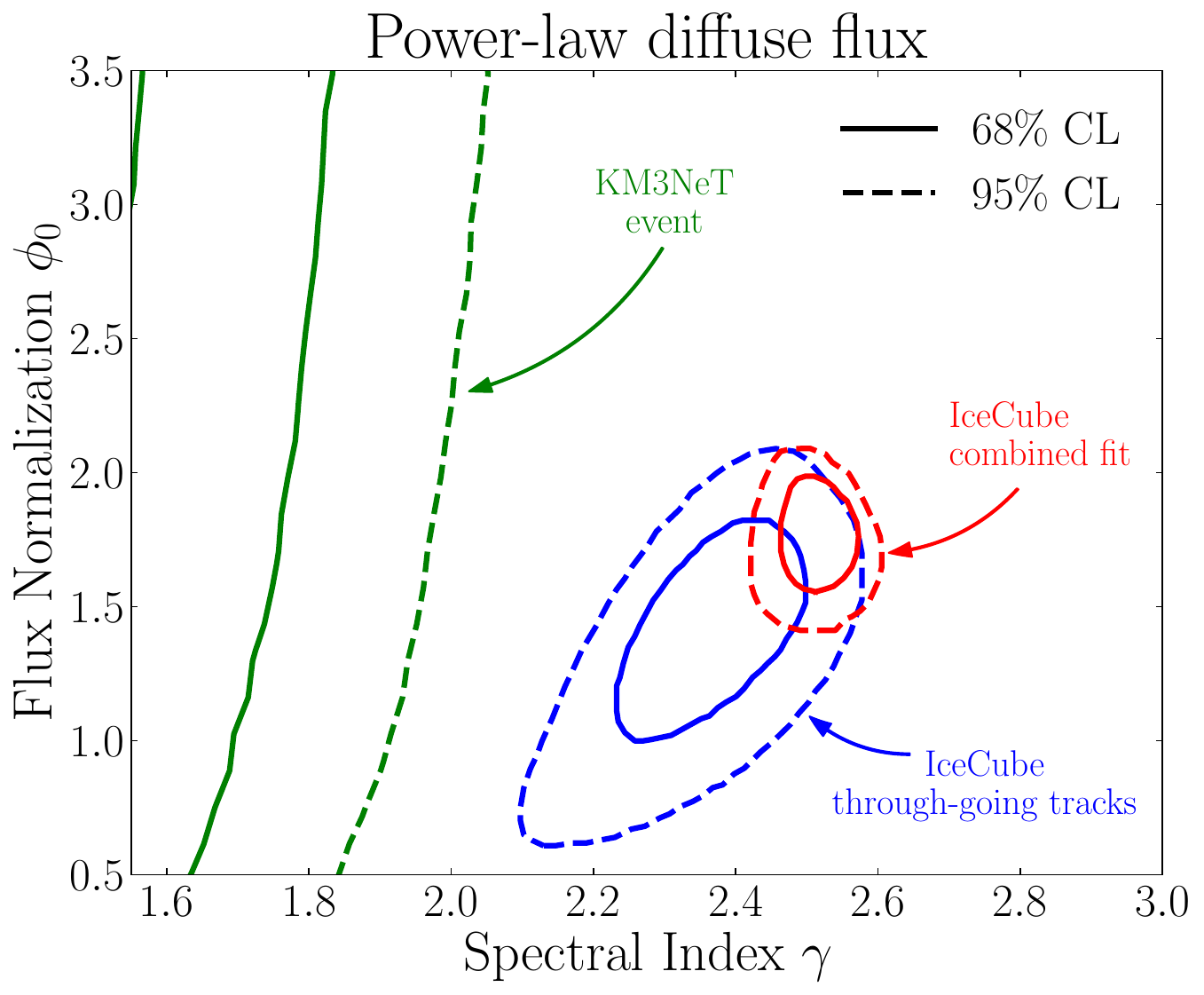}
    \includegraphics[width=0.48\linewidth]{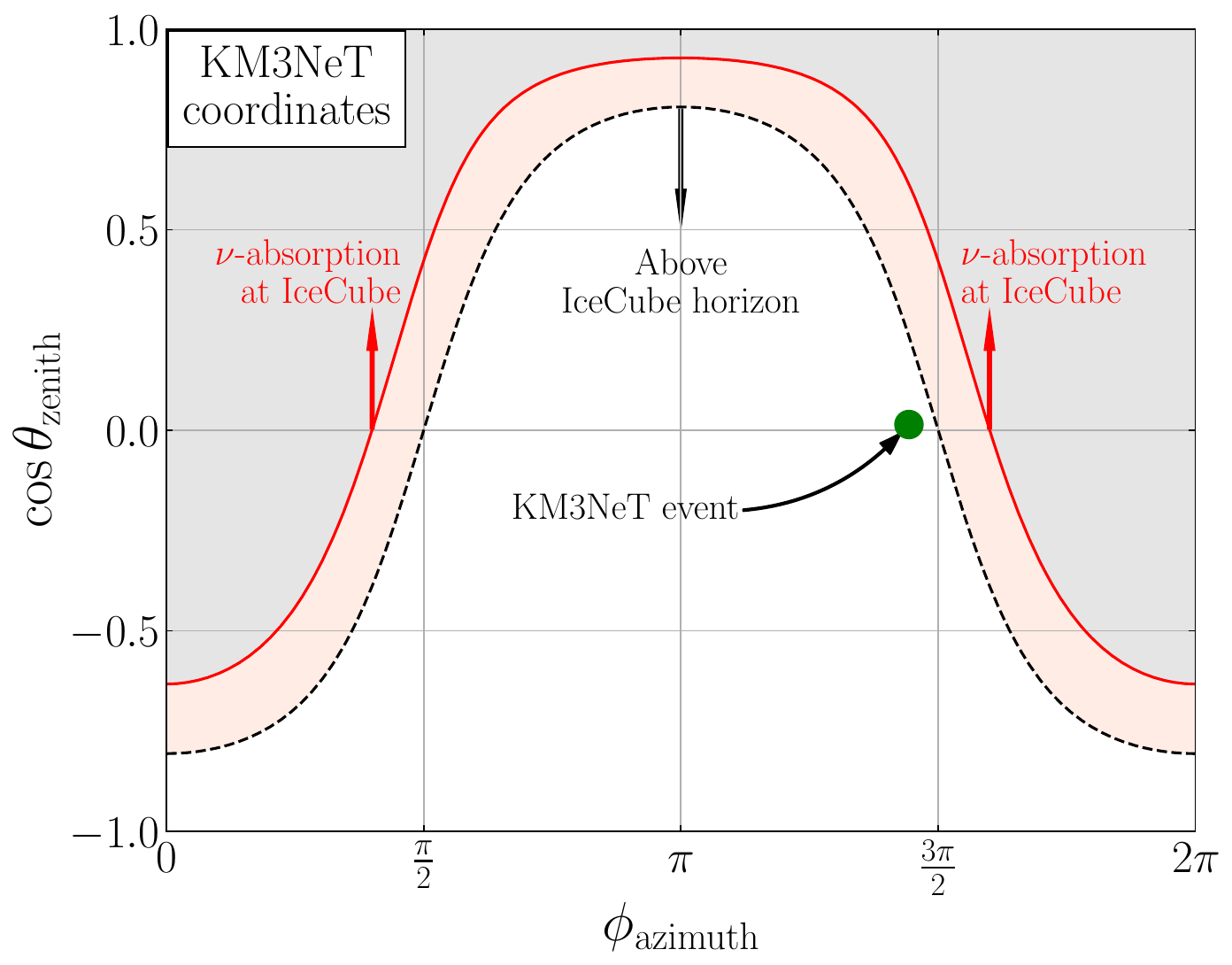} 
    \caption{Figures from~\cite{Li:2025tqf}. In the left panel, the green solid (dashed) contours show the 68\% (95\%) confidence levels from an independent analysis assuming a SPL diffuse flux and standalone KM3NeT/ARCA datataking. The IceCube SPL flux from~\cite{Naab:2023xcz} is shown in red and from~\cite{Abbasi:2021qfz} in blue. The right panel shows IceCube horizon (dashed black line) in KM3NeT local coordinates, with KM3-230213A indicated by the green dot.}
    \label{fig:tqf}
\end{figure}
In \cite{Li:2025tqf}, an independent analysis assessed the compatibility between KM3-230213A and published IceCube results. Using a toy detector model to simulate the number of $\npmtnohat$, as described in \cref{sec:enu}, three hypotheses were tested: diffuse power-law, cosmogenic, and point source neutrino fluxes. Keeping in mind that KM3NeT/ARCA was treated as a standalone detector, the results from a diffuse SPL fit is shown in green in the left panel of \cref{fig:tqf}. IceCube SPL measurements from~\cite{Naab:2023xcz} and~\cite{Abbasi:2021qfz} are shown in red and blue, respectively. With only one event, the green contour is largely unconstrained but prefer significantly different regions of the parameter space from IceCube. Even when the IceCube result from~\cite{Naab:2023xcz} is included as a gaussian prior, Ref.~\cite{Li:2025tqf} reports a \SI{3.5}{$\sigma$} tension between the two experiments under the SPL diffuse flux assumption.

In addition to the SPL, cosmogenic flux models, by which UHE neutrinos are produced by resonant CR interactions off the CMB, was tested in~\cite{Li:2025tqf} as well as~\cite{KM3NeT:2025vut}. The most favored cosmogenic model scenario by IceCube leads to an expectation of only \SI{5.7e-4}{events} over the ARCA21 exposure~\cite{Li:2025tqf}. By extending the UHECR source distribution up to redshift $z=6$ and introducing a small proton fraction at the higest energies, the expected number of ARCA21 events can be increased to $\mathcal{O}(10^{-2})$, thereby heightening the chances of such a detection while remaining largely consistent with IceCube upper limits~\cite{KM3NeT:2025vut}. However, if indeed ARCA21 expected to see such a relatively large number of events, the expectation rate in IceCube would be orders of magnitude larger. Even so, the rate is still low and non-observation could be attributed to statistics.

Finally, a point source or transient flux scenario was proposed as a method to reduce some of the tension between datasets~\cite{Li:2025tqf}. Although no significant sources have been associated with KM3-230213A, in part due to its relatively large directional uncertainty, its arrival direction does not preclude detection by IceCube due to down-going background or Earth absorption as shown in the right panel of \cref{fig:tqf}. Based on this, a steady point source hypothesis would be subject to the same exposure discrepancy as a diffuse flux scenario, and therefore still results in a \SI{2.9}{$\sigma$} tension. However, a transient source scenario could reduce the tension to \SI{2}{$\sigma$}, as it would reduce the impact of any livetime discrepancy between IceCube and ARCA21~\cite{Li:2025tqf}. This requires, however, a rather lucky scenario whereby a unique transient that exists as an outlier in any source population turned on at around the same time as ARCA21.

\section{Future prospects}
\label{sec:future}

The apparent detection of a potential EHE neutrino raises an intriguing question: is this event located on the falling high-energy tail of the astrophysical flux, or does it mark the onset of a cosmogenic component? In Sec.~\ref{sec:disc} we discussed several scenarios addressing the tension with IceCube results, as well as possible interpretations of the event’s origin. However, distinguishing between these possibilities will require both longer livetime and larger effective volumes, in order to perform a statistically meaningful test.

\begin{figure}[htbp]
  \centering
  \includegraphics[width=0.7\textwidth]{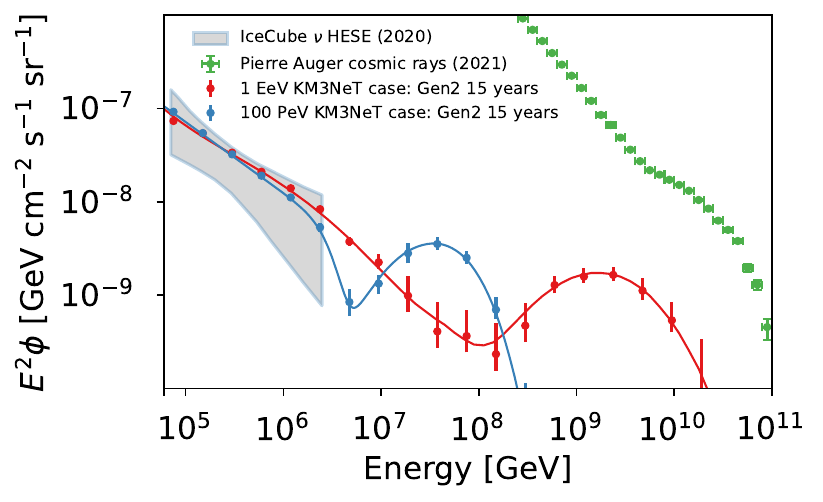}
    \caption{
    Projected diffuse astrophysical and cosmogenic neutrino measurements with IceCube-Gen2, assuming 15 years of combined optical and radio data taking~\cite{Muzio:2025gbr}. 
    The blue and red curves show best-fit models using combined IceCube, KM3NeT, and UHECR data from the Pierre Auger Observatory under a common-origin hypothesis. 
    The blue model assumes the KM3-230213A event has an energy of 100~PeV, yielding a best-fit with a $p\gamma$-induced neutrino bump at $\sim$30~PeV produced directly at the source. 
    The red model assumes the KM3NeT event has an energy of 1~EeV, with the best-fit requiring a cosmogenic origin by introducing a pure-proton subset of UHECRs. 
    Green points show the Pierre Auger UHECR flux~\cite{PierreAuger:2021hun}, and the grey band indicates IceCube’s High-Energy Starting Event (HESE) measurements~\cite{IceCube:2020wum}.
    }

  \label{fig:future_detectors_flux}
\end{figure}

Currently, IceCube has been operating in its full-detector configuration since 2011. 
Baikal-GVD~\cite{Zaborov:2024jny} and KM3NeT~\cite{Dornic2025_KM3NeTStatus}, both designed to reach instrumented volumes of order $1~\mathrm{km}^3$ comparable to IceCube, are under construction. 
In addition, several large-scale optical Cherenkov detectors have been proposed, including IceCube-Gen2~\cite{TDR}, P-ONE~\cite{P-ONE:2020ljt}, HUNT~\cite{Huang:2023mzt}, TRIDENT~\cite{TRIDENT:2022hql}, and NEON~\cite{Zhang:2024slv}, which would be distributed across the globe to provide complementary sky coverage. 
A summary of these experiments is presented in Tab.~\ref{tab:future}.

\begin{table}[t]
\centering
\footnotesize
\renewcommand{\arraystretch}{1.05}
\setlength{\tabcolsep}{3pt}

\begin{tabularx}{\textwidth}{l Y C C C C}
\toprule
Experiment & Site(s) & Volume [km$^3$] & Site depth [m] & Horizontal spacing [m] & Vertical spacing [m] \\
\midrule
IceCube     & South Pole & $\sim$1 & $\sim$2450 (ice) & 125 & 17 \\
IceCube-Gen2 & South Pole & $\sim$8 (proposed)& $\sim$2700 (ice) & 240 & 17 \\
KM3NeT--ARCA & Mediterranean Sea & $\sim$1 (target) & $\sim$3500 (sea) & 90--100 & 36--40 \\
Baikal--GVD  & Lake Baikal & $\sim$1 (target) & $\sim$1360 (lake) & 60 (intra); $\sim$300 (inter) & 15 \\
P-ONE        & Pacific Ocean & 1 (proposed) & 2660 (sea) & $\sim$100 & 50 \\
TRIDENT      & South China Sea & $\sim$7.5--8 (proposed) & 2800--3400 (sea) & 70--110 & 30--40 \\
NEON         & South China Sea & $\sim$10 (proposed) & 1700--3500 (sea) & 100 & 30 \\
HUNT         & South China Sea \& Baikal& $\sim$30 (proposed) & Baikal 1360; SCS 2560--3420 & $\sim$130$^{\dagger}$ & 36 \\
\bottomrule

\end{tabularx}
\caption{List of operational and planned Cherenkov neutrino telescopes. $^{\dagger}$ICRC\,2025 design study value, subject to optimisation.}
\label{tab:future}
\end{table}

With IceCube-Gen2, the instrumented volume will be increased to about eight times that of IceCube, significantly accelerating the detection of the highest-energy neutrinos. Using both through-going tracks and events with interaction vertices contained within the Gen2 volume, we performed simulation studies to compute the piecewise-unfolded spectrum under two signal hypotheses: (i) the KM3-230223A event is of astrophysical origin, with an energy of order 100~PeV emitted directly from the source, or (ii) it originates from the GZK cosmogenic component produced by ultra-high-energy cosmic rays interacting with cosmic microwave background photons (red). Both models were calculated following a global diffuse fit~\cite{Muzio:2025gbr} including data from IceCube, KM3NeT and Pierre Auger as introduced in Sec.\ref{sec:disc}. For 15 years of Gen2 data taking, the shown error bars reflect statistical uncertainties only. Detector systematics are expected to remain small thanks to a decade of studies on ice properties and the upcoming IceCube Upgrade. Atmospheric neutrino uncertainties are negligible at these energies. Under these assumptions, a clear separation between the two hypotheses emerges.

Beyond optical Cherenkov techniques, alternative detection methods are being developed to enhance sensitivity to EHE neutrinos. For instance, the space-based POEMMA~\cite{POEMMA:2020ykm} mission (Probe of Extreme Multi-Messenger Astrophysics) is designed to observe Cherenkov and fluorescence photons produced by upward-going air showers initiated by $\nu_\tau$ interactions. Radio-based detection methods, which exploit the Askaryan effect in dense media like ice, are also gaining momentum. Notable examples include the ARA~\cite{ARA:2019wcf} in Antarctica and RNO-G~\cite{RNO-G:2020rmc} in Greenland.

Other promising approaches focus on detecting radio signals from tau neutrinos interacting in mountainous terrain. These include the proposed GRAND~\cite{GRAND:2018iaj}, BEACON~\cite{Wissel:2020sec} and TAMBO~\cite{Romero-Wolf:2020pzh} experiments. The ANITA~\cite{Gorham:2019guw} balloon experiment has already conducted several flights to search for Askaryan radio pulses reflected off the ice surface, proceeded by PUEO~\cite{PUEO:2020bnn}. Lunar-based detection concepts such as NuMoon~\cite{Buitink:2008bc} aim to observe Askaryan signals from neutrino interactions within the Moon's regolith.

Additional techniques include TRINITY~\cite{Brown:2021ane}, a proposed experiment targeting Cherenkov photons from air showers initiated by emerging $\nu_\tau$ near mountains, and the Radar Echo Telescope (RET)~\cite{Prohira:2019glh}, which seeks to detect radar reflections from particle cascades in ice.

The growing number of ongoing and proposed EHE neutrino experiments reflects a shared scientific goal: to uncover the origins of ultra-high-energy cosmic particles and to probe fundamental physics at energy scales far beyond the reach of terrestrial accelerators.

\begin{figure}[htbp]
  \centering
  \includegraphics[width=0.8\textwidth]{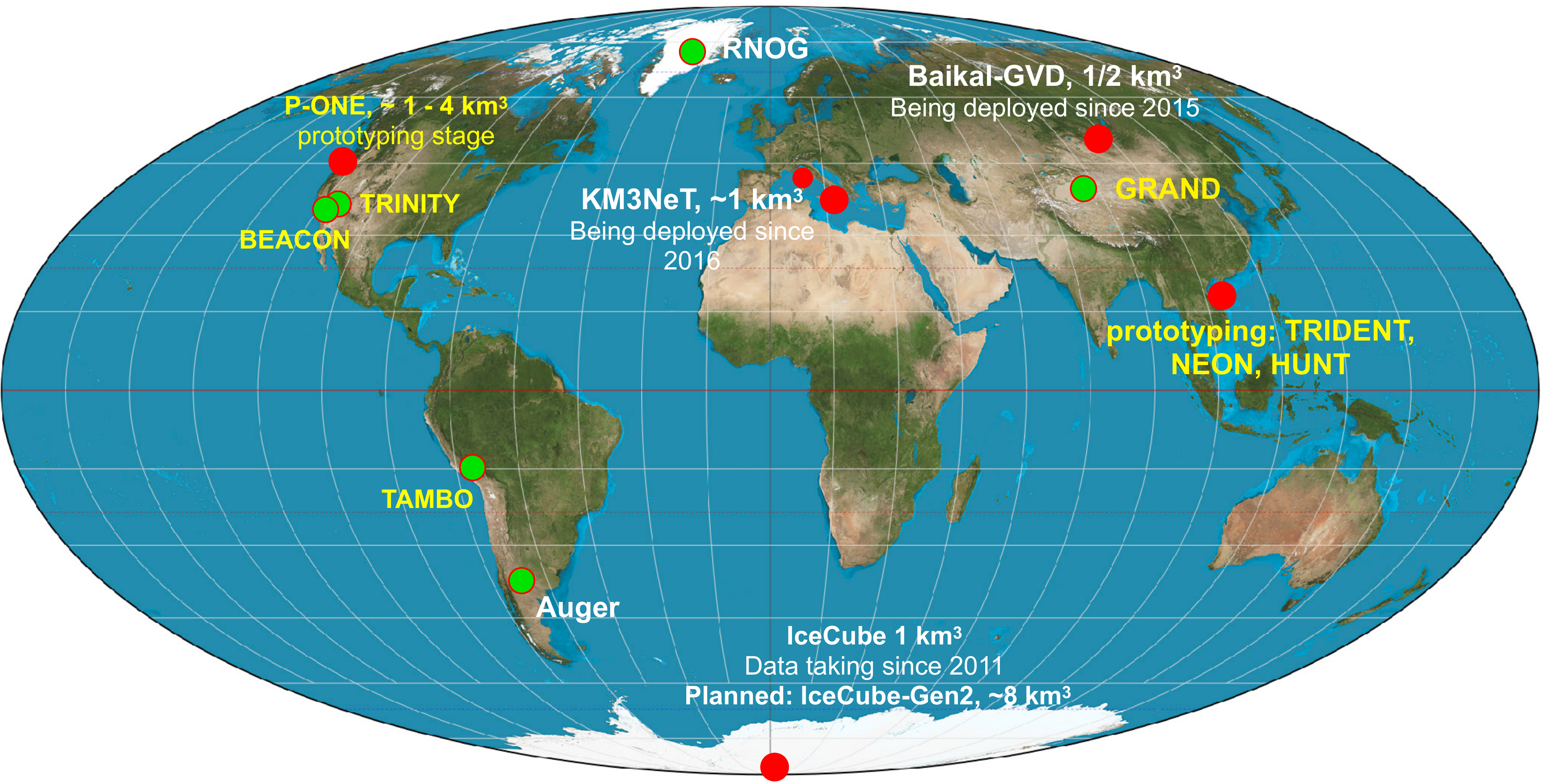}
  \caption{Existing and proposed sites of future neutrino telescopes. Detectors using the in-ice or in-water Cherenkov technique are shown in red. Experiments aiming for the detection of EHE neutrinos via the Earth-skimming technique or the Askaryan effect in ice are shown in green.}
  \label{fig:future_detectors}
\end{figure}

The locations of current and proposed ground-based neutrino observatories targeting the highest energies are shown in Fig.~\ref{fig:future_detectors}. A geographically distributed network of detectors provides complementary sky coverage, mitigating the issue of Earth absorption for the highest-energy neutrinos during propagation. Such a configuration also enables real-time detection and follow-up observations by covering a larger fraction of the sky. Furthermore, coordinated operation between observatories allows for spatial--temporal correlation studies, enhancing the search for neutrino multiplets.

Building on this global perspective, Fig.~\ref{fig:future_detectors_flux} compares measured diffuse neutrino fluxes from IceCube, Auger, and KM3NeT with the projected sensitivities of proposed future non-optical experiments. The only exception in this comparison is IceCube-Gen2, which will feature both an optical array and a large-scale radio array. At energies above $\sim 100$~PeV, the sensitivity of Gen2 is dominated by its radio component, consisting of a $\mathcal{O}(10^2)\,\mathrm{km}^2$ array of antennas deployed deep in the Antarctic ice.

\begin{figure}[htbp]
  \centering  \includegraphics[width=0.9\textwidth]{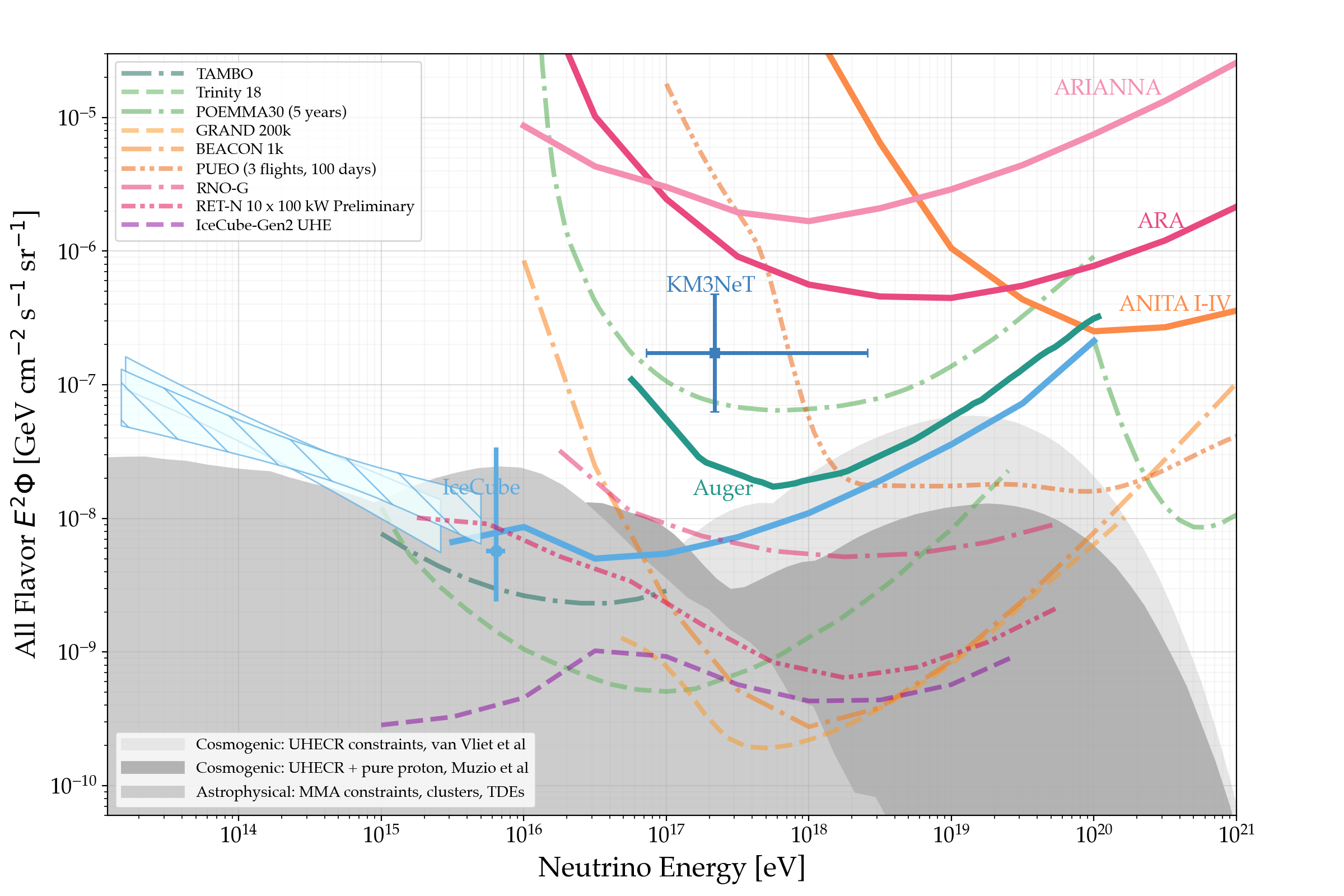}
    \caption{
    Overview of present measurements and projected sensitivities for the diffuse neutrino flux across a broad energy range. 
    This figure is adapted from~\cite{Ackermann:2022rqc}. 
    The hatched blue band represents IceCube’s astrophysical $\nu_\mu$ flux from through-going tracks~\cite{Abbasi:2021qfz}, while the solid blue band corresponds to the $\nu_e+\nu_\tau$ flux derived from cascade-like events~\cite{IceCube:2020acn}. 
    The light blue marker indicates the flux level implied by the Glashow resonance candidate~\cite{IceCube:2021rpz}, and the dark blue marker shows the KM3NeT event KM3-230213A~\cite{KM3NeT:2025npi}. 
    Solid curves denote upper limits from existing high-energy neutrino searches by the Pierre Auger Observatory~\cite{PierreAuger:2023pjg}, ARA~\cite{ARA:2019wcf}, ARIANNA~\cite{Anker:2019rzo}, ANITA I–IV~\cite{Gorham:2019guw}, and IceCube~\cite{Meier:2025nql}. 
    Dashed curves show the anticipated sensitivities of several proposed or under-construction experiments, including GRAND with 200,000 antennas~\cite{GRAND:2018iaj}, BEACON with 1000 stations~\cite{Wissel:2020sec}, TAMBO with 22,000 detectors~\cite{Romero-Wolf:2020pzh}, Trinity with 18 telescopes (updated from~\cite{Brown:2021ane}), RET-N with 10 stations each using a 100~kW transmitter~\cite{Prohira:2019glh}, POEMMA30~\cite{POEMMA:2020ykm}, RNO-G~\cite{RNO-G:2020rmc,RNO-G:2021hfx}, and PUEO~\cite{PUEO:2020bnn}. 
    All sensitivities are shown as differential 90\%~C.L. limits for the all-flavour flux, computed in decade-wide energy bins for a nominal ten-year exposure unless noted otherwise. 
    Selected model predictions for cosmogenic neutrinos~\cite{vanVliet:2019nse, Muzio:2021zud} and for astrophysical source scenarios~\cite{Muzio:2021zud, Fang:2017zjf, Biehl:2017hnb} are overlaid for comparison.
    }
    
  \label{fig:snowmass}
\end{figure}

These improved sensitivities will enable a wide range of physics opportunities. 
In the ultra-high-energy regime, they will allow a decisive test of whether the diffuse neutrino flux continues to fall with energy as observed by IceCube, or if an additional cosmogenic component emerges due to ultra-high-energy cosmic rays interacting with background photon fields. 
Events such as KM3-230223A highlight the importance of expanding detection capabilities: with greater exposure and complementary detection techniques, future observatories will improve the chances of identifying neutrino point sources and performing time-dependent searches for transient phenomena such as gamma-ray bursts or tidal disruption events. 
Furthermore, the broad energy coverage and enhanced statistics will facilitate precision measurements of the neutrino cross section and flavour composition~\cite{Ackermann:2022rqc}, providing knowledge on both astrophysical acceleration mechanisms and neutrino properties over energy scales far beyond the reach of terrestrial accelerators.

\clearpage
\section*{Acknowledgement}
The authors thank M.~Lamoureux for providing helpful clarifications related to KM3NeT, N.~M.~Binte~Amin for providing the IceTop veto event visualization, N.~Kamp and A.~Schneider for guidance with the SIREN framework, and D.~Chirkin and A.~Karle for many useful discussions related to energy estimation, event reconstruction and the interesting physics pertinent to all neutrino telescopes. The authors were supported in part by NSF grant PHY-2513077, IIS-2435532 and by the University of Wisconsin Research Committee with funds granted by the Wisconsin Alumni Research Foundation.

\printbibliography

\end{document}